\documentclass[runningheads]{llncs}
\usepackage{isabelle,isabellesym}

\usepackage[ruled,vlined]{algorithm2e}

\usepackage {tikz}
\usetikzlibrary {positioning}
\usetikzlibrary{arrows.meta,positioning,calc}
\usetikzlibrary{graphs,graphs.standard,quotes}

\usepackage{multicol}
\usepackage{amsmath}
\usepackage{amssymb}

\usepackage{pdfsetup}

\isabellestyle{it}

\begin{document}

\title{Formalizing Graph Trail Properties in Isabelle/HOL}
\author{Laura Kov{\'a}cs \and Hanna Lachnitt \and Stefan Szeider}
\authorrunning{L. Kov{\'a}cs et al.}
\institute{TU Wien, Vienna, Austria\\
	\email{\{laura.kovacs,hanna.lachnitt,stefan.szeider\}@tuwien.ac.at}}
\maketitle

\begin{abstract}
	We describe a dataset expressing and proving properties of graph trails, using Isabelle/HOL.
	We formalize the reasoning about strictly increasing and decreasing trails, using weights over edges,
	and prove lower bounds over the length of trails in weighted graphs. We do so by extending the graph theory library of Isabelle/HOL with an algorithm computing the length of a longest strictly decreasing graph trail starting from a vertex for a given weight distribution, and prove that any decreasing trail is also an increasing one.
\end{abstract}


%
\begin{isabellebody}%
\setisabellecontext{Ordered{\isacharunderscore}Trail}%
\isadelimtheory
\endisadelimtheory
\isatagtheory
\endisatagtheory
{\isafoldtheory}%
\isadelimtheory
\endisadelimtheory
\isadelimdocument
\endisadelimdocument
\isatagdocument
\isamarkupsection{Introduction%
}
\isamarkuptrue%
\endisatagdocument
{\isafolddocument}%
\isadelimdocument
\endisadelimdocument
\begin{isamarkuptext}%
The problem of finding a longest trail with strictly increasing or strictly decreasing weights in
an edge-weighted graph is an interesting graph theoretic problem~\cite{graham1973increasing,calderbank1984increasing,yuster2001large,de2015increasing}, 
with potential
applications to scheduling and cost distribution in traffic planning and routing~\cite{byron}.
In this paper, we formalize and automate the reasoning about
strictly increasing and strictly decreasing trail properties by developing an extendable flexible library in the proof assistant Isabelle/HOL~\cite{nipkow2002isabelle}.

As a motivating example consider the following (undirected) graph $K_4$, where each edge is annotated 
with a different integer-valued weight ranging from $1, \ldots, 6$:

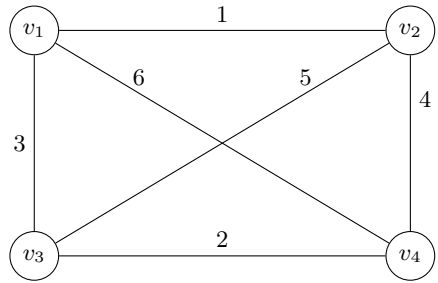
\begin{figure}
\centering
		\begin{tikzpicture}
		\node[draw,circle] at (0, 0) (a) {$v_1$};
		\node[draw,circle] at (5, 0) (b) {$v_2$};
	
		\node[draw,circle] at (0, -3) (c) {$v_3$};
		\node[draw,circle] at (5, -3) (d) {$v_4$};
		
		\draw[] (a) -- (b)  node[above,midway] {$1$};
		\draw[] (a) -- (c)  node[midway, left] {$3$};
		\draw[] (a) -- (d)  node[near start, above] {$6$};
	  
    \draw[] (b) -- (c)  node[near start, above] {$5$};
		\draw[] (b) -- (d)  node[near start, right] {$4$};
		
		\draw[] (c) -- (d)  node[midway, above] {$2$};
		\end{tikzpicture}
  \caption{Example graph $\protect K_4$}\label{example:K4}
\end{figure}

\noindent When considering $K_4$, the question we address in this paper is whether $K_4$ has a 
strictly decreasing trail of length $k\geq 1$. A trail is a sequence of distinct edges $(e_1,\ldots,e_k)$, $e_i \in E$ such that
there exists a corresponding sequence of vertices $(v_0,...,v_k)$ where $e_i = v_{i-1}v_i$. 
A strictly-ordered trail is a trail where the edge weights of $(e_1,\ldots,e_k)$ are either strictly increasing or strictly decreasing.
Our work provides a formally verified algorithm 
computing such strictly-ordered trails. Note that there is a decreasing trail in $K_4$ starting at vertex $v_3$, 
with trail length 3; namely $(v_3v_2; v_2v_4; v_4v_3)$ is such a trail, with each edge in the trail having 
a higher weight than its consecutive edge in the trail. Similarly, $K_4$ has decreasing trails of 
length 3 starting from $v_1$, $v_2$, and $v_4$ respectively. A natural question to ask,
which we address in this paper, is whether it is possible to construct a graph such that the 
 constructed graph has 4 vertices and 5 edges, and no vertex is the starting node of a trail of 
length 3? We answer this question negatively, in an even more general setting,  not restricted to 4 
vertices and 5 edges. Similarly to the theoretical results of~\cite{graham1973increasing}, we show 
that, given a graph $G$ with $n$ vertices and $q$ edges, there is always a strictly decreasing trail of length at least
$2 \cdot \lfloor\frac{q}{n}\rfloor$. While such a graph theoretical result has already been announced 
\cite{graham1973increasing}, in this paper we formalize the results in Isabelle/HOL and construct 
a Isabelle/HOL-verified algorithm computing strictly decreasing trails of length $k$, whenever such trails exist.

Let us note that proving that a
 graph $G$ with $n$ vertices and $q$ edges has/does not have decreasing trails is possible for small 
$n$, using automated reasoning engines such as Vampire~\cite{Vampire13} and Z3~\cite{de2008z3}. 
One can restrict the weights to the integers $1,..,q$ and since $q \le {n \choose 2}$ there is a 
finite number of possibilities for each $n$. 
Nevertheless, the limit of such an undertaking is
reached soon. On our machine\footnote{standard laptop with
1.7 GHz Dual-Core Intel Core i5 and  8 GB 1600 MHz memory} even for $n$ = 7, both Vampire and Z3 fail 
proving the existence of strictly decreasing trails, using a 1 hour time limit. This
is due to the fact that every combination of edge weights and starting nodes is
tested to be a solution. Thus, the provers are not able to contribute to the process of finding an 
effective proof of the statement. Even for relatively small numbers $n$, our experiments show that 
state-of-the-art automated  provers are not able to prove whether weighted graphs have a strictly decreasing trail 
of a certain length.

We also note that this limitation goes beyond automated provers. In the Isabelle proof assistant, 
proving that a complete graph with 3 vertices, i.e. $K_3$,  will
always contain a strictly decreasing trail of length 3 is quite exhaustive, as it requires reasoning about 3! = 6 possibilities
for a distribution of a weight function $w$ and then manually constructing concrete trails:

\begin{center}
\isa{w{\isacharparenleft}v\isactrlsub {\isadigit{1}}{\isacharcomma}v\isactrlsub {\isadigit{2}}{\isacharparenright}\ {\isacharequal}\ {\isadigit{2}}\ {\isasymand}\ w{\isacharparenleft}v\isactrlsub {\isadigit{2}}{\isacharcomma}v\isactrlsub {\isadigit{3}}{\isacharparenright}\ {\isacharequal}\ {\isadigit{1}}\ {\isasymand}\ w{\isacharparenleft}v\isactrlsub {\isadigit{3}}{\isacharcomma}v\isactrlsub {\isadigit{1}}{\isacharparenright}\ {\isacharequal}\ {\isadigit{3}}} 

\isa{{\isasymlongrightarrow}\ incTrail\ K\isactrlsub {\isadigit{3}}\ w\ {\isacharbrackleft}{\isacharparenleft}v\isactrlsub {\isadigit{3}}{\isacharcomma}v\isactrlsub {\isadigit{2}}{\isacharparenright}{\isacharcomma}{\isacharparenleft}v\isactrlsub {\isadigit{2}}{\isacharcomma}v\isactrlsub {\isadigit{1}}{\isacharparenright}{\isacharcomma}{\isacharparenleft}v\isactrlsub {\isadigit{1}}{\isacharcomma}v\isactrlsub {\isadigit{3}}{\isacharparenright}{\isacharbrackright}}
\end{center}

Based on such limitations of automative and interactive provers, in this paper we aim at formalizing 
and proving existence of trails of length $n$, where $n\geq 1$ is a symbolic constant. As such, 
proving for example that graphs have trails of length $4$, for a concrete $n$,  become instances 
of our approach. To this end, we build upon existing works in this area. In particular, the first 
to raise the question of the minimum length of strictly increasing
trails of arbitrary graphs  were Chv\'atal and Koml\'os~\cite{chvatal1970some}. Subsequently, 
Graham and Kletman~\cite{graham1973increasing} proved that the lower bound of the length of increasing trails 
is given by $2 \cdot \lfloor\frac{q}{n}\rfloor$, as also mentioned above. In our work, we formalize and verify 
such results in Isabelle/HOL. Yet, our work is not a straightforward adaptation and formalization   
of Graham and Kletman's proof~\cite{graham1973increasing}. Rather, we focus on decreasing trails instead of increasing trails 
and give an algorithm computing longest decreasing trails of a given graph (Algorithm \ref{algo:FindLongestTrail}).  
By formalizing Algorithm \ref{algo:FindLongestTrail} in Isabelle/HOL, we also formally verify the correctness 
of the trails computed by our approach. Moreover, we prove that any strictly decreasing trail is also an strictly increasing 
one, allowing this way to use our formalization in Isabelle/HOL also to formalize results of Graham and Kletman~\cite{graham1973increasing}.

\noindent\paragraph{\bf Contributions.} This paper brings the following contributions.
\begin{description}
\item[(1)]
We formalize strictly increasing trails and provide basic lemmas about their
properties. We improve results of \cite{graham1973increasing} by giving a precise bound on the increase of trail length.
\item[(2)] We formalize strictly decreasing trails, in addition to the increasing trail setting of~\cite{graham1973increasing}. 
We prove the duality between strictly increasing and strictly decreasing trails, that is, any such decreasing trail is an increasing one, and vice versa. 
Thanks to these extensions, unlike \cite{graham1973increasing},  we give a constructive proof of the existence of strictly ordered trails (Lemma \ref{lemma:sum}). 
\item[(3)] We design an algorithm computing longest ordered trails (Algorithm \ref{algo:FindLongestTrail}), and formally verify  its correctness in Isabelle/HOL.
We extract our algorithm to Haskell program code using Isabelle's program extraction tool. Thus, we obtain a fully verified algorithm to compute the length
of strictly-ordered trails in any given graph and weight distribution.
\item[(4)] We
verify the lower bound on
the minimum length of strictly decreasing trails of arbitrary graphs, and of complete graphs in particular.
\item[(5)] We build upon the Graph-Theory library by Noschinski~\cite{Graph_Theory-AFP},  that is part of the
Archive of Formal Proofs (AFP) and already includes many results on walks and
general properties of graphs. We introduce the digital dataset {\it v} formalizing properties of graph trails. Our dataset  consists of
$\sim$2000 lines of Isabelle code and it took about one month for one person to finish. As far as we know this is the first formalization of
ordered trails in a proof assistant.
\end{description}

This paper was generated from Isabelle/HOL source code using Isabelle's document preparation tool 
and is therefore fully verified. The source code is available
online at \url{https://github.com/Lachnitt/Ordered_Trail}. The rest of the paper is organized as follows.
Section 2 recalls basic terminology and properties from graph theory. 
We prove lower bounds on strictly increasing/decreasing trails in Section 3. We describe our Isabelle/HOL 
formalization in Isabelle/HOL in Section 4. We discuss further directions in Section 5 and conclude our paper with Section 6.%
\end{isamarkuptext}\isamarkuptrue%
\isadelimdocument
\endisadelimdocument
\isatagdocument
\isamarkupsection{Preliminaries%
}
\isamarkuptrue%
\endisatagdocument
{\isafolddocument}%
\isadelimdocument
\endisadelimdocument
\begin{isamarkuptext}%
\label{section:Prelim}
We briefly recapitulate the basic notions of graph theory. A {\em graph} $G = (V,E)$ consists of
a set $V$ of {\em vertices} and a set $E \subseteq V \times V$ of {\em edges}. A graph is undirected 
if $(v_1,v_2)\in E$ implies that also $(v_2,v_1)\in E$. A graph is {\em complete}
 if every pair of vertices is connected by an edge. A graph is {\em loopfree} or {\em simple} if there are no edges $(x,x)\in E$ and 
{\em finite} if the number of vertices $|V|$ is finite. Finally, we call 
a graph $G'=(V',E')$ a {\em subgraph} of $G = (V,E) $ if $V' \subseteq V$ and $E' \subseteq E$.

If a graph is equipped with a
 weight function $w: E \rightarrow \mathbb{R}$ that maps edges to real numbers, it is called 
 an {\em edge-weighted graph}. In the following, whenever a graph is mentioned it is implicitly assumed
that this graph comes equipped with a weight function. A vertex labelling is a function $L: V \rightarrow \mathbb{N}$.

A {\em trail of length k} in a graph $G = (V,E)$ is a sequence $(e_1,\ldots,e_k)$, $e_i \in E$, of distinct edges such that
 there exists a corresponding sequence of vertices $(v_0,...,v_k)$ where $e_i = v_{i-1}v_i$. 
A {\em strictly decreasing trail} in an edge-weighted graph $G = (V,E)$ with weight function $w$
is a trail such that $w (e_i) > w (e_{i+1})$. Likewise, a {\em strictly increasing trail} is a trail such that $w (e_i) < w (e_{i+1})$.
A trail is {\em strictly-ordered} if it is strictly increasing or strictly decreasing.

We will denote the length of a longest strictly increasing trail with $P_i(w,G)$. Likewise we will denote the length
of a longest strictly decreasing trail with $P_d(w,G)$. In any undirected graph, it holds that $P_i(w,G) = P_d(w,G)$, 
a result that we will formally verify in Section \ref{section:trails}. 

Let $f_i(n) = \min_n P_i(w,K_n)$ denote the minimum length of an strictly increasing trail that must exist in 
the complete graph with $n$ vertices. Likewise, $f_d(n) = \min_n P_d(w,K_n)$ in the case that we consider 
strictly decreasing trails.%
\end{isamarkuptext}\isamarkuptrue%
\isadelimdocument
\endisadelimdocument
\isatagdocument
\isamarkupsection{Lower Bounds on Increasing and Decreasing Trails in Weighted Graphs%
}
\isamarkuptrue%
\endisatagdocument
{\isafolddocument}%
\isadelimdocument
\endisadelimdocument
\begin{isamarkuptext}%
\label{section:symbolicProof}
The proof introduced in the following is based on similar ideas as in \cite{graham1973increasing}. 
However, we diverge from \cite{graham1973increasing} in several aspects. Firstly, we consider strictly decreasing instead of strictly increasing trails,
reducing the complexity of the automated proof (see Section \ref{section:Formalization}). 
Moreover, we add tighter bounds than necessary to give a fully constructive proof in terms of an algorithm for computing
the length of these trails (see Section \ref{section:localeSurjective}). We discuss this further at the end of the section.

We start by introducing the notion of a weighted subgraph and then we built on that by specifying a family of labelling functions:

\begin{definition}[Weighted Subgraph] \label{def:weightedSubgraph}
	Let $G=(V,E)$ be a graph with weight function $w:E\rightarrow \{1,\ldots,q\}$ where $|E| = q$.
	For each $i\in \{0,...,q\}$ define a weighted subgraph $G^i = (V,E^i)$ such that $e\in E^i$ iff $w(e)\in \{1,...,i\}$. 
  That is, $G^i$ contains only edges labelled with weights $\le i$.
\end{definition}

\begin{definition}[Labelling Function]\label{def:Labelling}
	For each $G^i=(V,E^i)$, $n = |V|$  we define $L^i:V \{1,\ldots,\frac{n(n-1)}{2}\}$ a labelling function such 
that $L^i(v)$ is the length of a longest strictly decreasing trail starting at vertex v using only edges in $E^i$.
\end{definition}

\noindent In Figure \ref{example:G5} the example graph from Figure \ref{example:K4} is revisited to 
illustrate these definitions. We need to prove the following property.

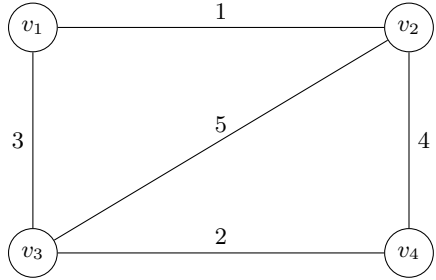
\begin{figure}
\centering

	\begin{multicols}{2}[\columnsep2em] 

	\begin{tikzpicture}
	\node[draw,circle] at (0, 0)   (a) {$v_1$};
	\node[draw,circle] at (5, 0)   (b) {$v_2$};
	
	\node[draw,circle] at (0, -3)  (c)     {$v_3$};
	\node[draw,circle] at (5, -3)  (d)     {$v_4$};
	
	\draw[] (a) -- (b)  node[above,midway] {$1$};
  \draw[] (a) -- (c)  node[midway, left] {$3$};
	\draw[] (b) -- (d)  node[midway, right] {$4$};
	
	\draw[] (b) -- (c)  node[midway, above] {$5$};
	\draw[] (c) -- (d)  node[midway, above] {$2$};
	
	\end{tikzpicture}\\
		\columnbreak
		\vspace{7cm}
		\ \\
		Decreasing trails from $v_3$ are: 
		
		$v_3-v_4$,  
		
		$v_3-v_1-v_2$, 
		
		$v_3-v_2-v_1$,
		
		$v_3-v_2-v_4-v3$
		
		Therefore, $L^5(v_3) = 3$.\\ \ \\

		Decreasing trails from $v_1$ are:
		
		$v_1-v_2$
		
		$v_1-v_3-v_4$
		
		Therefore, $L^5(v_1) = 2$.
	\end{multicols}
	
  \caption{Graph $G^5$ with labelling function $L^5$}\label{example:G5}
\end{figure}

\begin{lemma}\label{lemma:sum}
If $i < q$, then $\sum_{v\in V} L^{i+1}(v) \ge \sum_{v\in V} L^{i}(v)+2$.
\end{lemma}
\vspace{-1em}
\begin{proof}
Let $e$ be the edge labelled with $i+1$ and denote its endpoints with $u_1$ and $u_2$. It holds that $E^i \cup \{e\} = E^{i+1}$, 
therefore the graph $G^{i+1}$ is $G^i$ with the additional edge $e$. As $w(e') < w(e) $, for all $e' \in E^i$ we have 
 $L^{i+1}(v) = L^i(v)$ for all $v\in V$ with $u_1\neq v, u_2\neq v$. It also holds that $L^{i+1}(u_1) = \max(L^i(u_2)+1,L^i(u_1))$ 
because either that longest trail from $u_1$ can be prolonged with edge $e$ ($i+1$ will be greater than the weight of the first edge 
in this trail by construction of $L^{i+1}$) or there is already a longer trail starting from $u_1$ not using e. 
We derive $L^{i+1}(u_2) = \max(L^i(u_1)+1,L^i(u_2))$ based on a similar reasoning. See Figure \ref{figure:cases} for an illustration. 

Note that $L^{i+1}(v) = L^i(v)$ for $v\in V \setminus \{u_1,u_2\}$, because no edge incident to these vertices was added and
a trail starting from them cannot be prolonged since the new edge has bigger weight than any edge in such a 
trail.

If $L(u_1)=L(u_2)$, then $L^{i+1}(u_1) = L^i(u_1) + 1$ and $L^{i+1}(u_2) = L^i(u_2)+1$ and thus the sum increases exactly by 2. 
If $L(u_1)>L(u_2)$ then $L^{i+1}(u_2) = L^i(u_1) +1 \ge L^i(u_2)+2$, otherwise $L^{i+1}(u_1) = L^i(u_2) +1 \ge L^i(u_1)+2$. Thus, 

\begin{flalign*}
\sum_{v\in V} L^{i+1}(v) &~ = ~& \sum_{v\in (V-\{u_1,u_2\})} L^{i+1}(v)+L^{i+1}(u_1)+L^{i+1}(u_2)&&\\
&~\ge~ & \sum_{v\in (V-\{u_1,u_2\})} L^{i+1}(v)+L^i(u_1)+L^i(u_2)+2&&\\
&~=~ &  \sum_{v\in V} L^{i}(v)+2.&&
\end{flalign*}\qed
\end{proof}

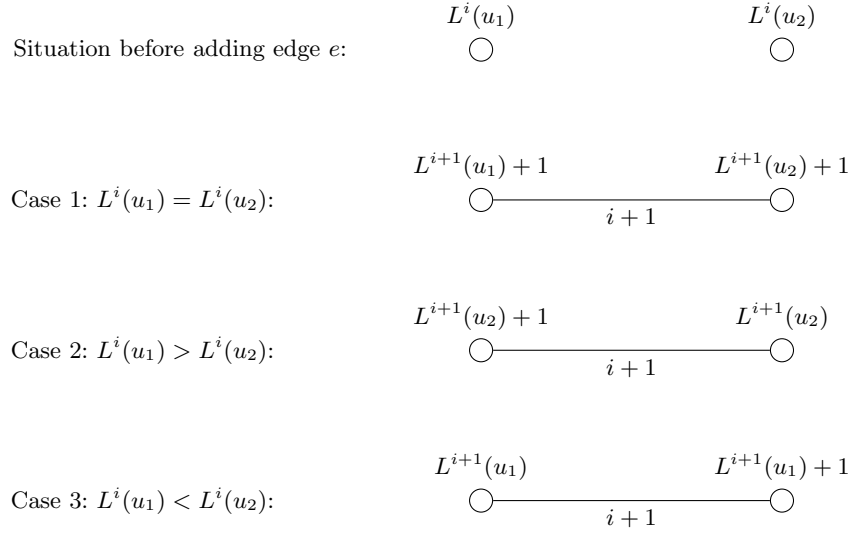
\begin{figure}[h]
	\centering
	\begin{tikzpicture}
	\node at (-4, 6)  {Situation before adding edge $e$: }; 
	\node[draw,circle, label = $L^i(u_1)$] at (0, 6)   (a) {};
	\node[draw,circle, label = $L^i(u_2)$] at (4, 6)   (b) {};
	
	\node at (-4.5, 4)  {Case 1: $L^i(u_1) = L^i(u_2)$: }; 
	\node[draw,circle, label = $L^{i+1}(u_1)+1$] at (0, 4)   (c) {};
	\node[draw,circle, label = $L^{i+1}(u_2)+1$] at (4, 4)   (d) {};			
	\draw[] (c) -- (d)  node[below,midway] {$i+1$};
	
	\node at (-4.5, 2)  {Case 2: $L^i(u_1) > L^i(u_2)$:  }; 
	\node[draw,circle, label = $L^{i+1}(u_2)+1$] at (0, 2)   (c) {};
	\node[draw,circle, label = $L^{i+1}(u_2)$] at (4, 2)   (d) {};			
	\draw[] (c) -- (d)  node[below,midway] {$i+1$};
	
	\node at (-4.5, 0)  {Case 3: $L^i(u_1) < L^i(u_2)$:}; 
	\node[draw,circle, label = $L^{i+1}(u_1)$] at (0, 0)   (c) {};
	\node[draw,circle, label = $L^{i+1}(u_1)+1$] at (4, 0)   (d) {};			
	\draw[] (c) -- (d)  node[below,midway] {$i+1$};
	\end{tikzpicture}
	\caption{Case distinction when adding edge $e$ in Lemma \ref{lemma:sum}}\label{figure:cases}
\end{figure}

\noindent Note that the proof of Lemma \ref{lemma:sum} is constructive, yielding the Algorithm \ref{algo:FindLongestTrail} for computing
longest strictly decreasing trails. Function $findEndpoints$ searches for an edge in a graph $G$ by its weight $i$ and returns
both endpoints. Function $findMax$ returns the maximum value of the array $L$. 

\begin{algorithm}[H]
	\SetAlgoLined
	
	\For{$v\in V$}{$L(v):=0$}
	\For{$i=1; i<|E|; i++$}{
		$(u,v) = findEndpoints(G,i)$\;
		$temp = L(u)$\;
		$L(u) = \max(L(v)+1,L(u))$ \;
		$L(v) = \max(temp+1,L(v))$ \;	
	}

	return findMax(L)\;

\caption{Find Longest Strictly Decreasing Trail}\label{algo:FindLongestTrail}
\end{algorithm}

\begin{lemma}$\sum_{v\in V} L^{q}(v) \ge 2q$. \end{lemma}

\begin{proof}
We proceed by induction, using the property $\sum_{v\in V} L^{i+1}(v) \ge \sum_{v\in V} L^{i}(v)+2$ from Lemma \ref{lemma:sum}. For the induction base note that $\sum_{v\in V} L^{0}(v) = 0$ 
because $G^0$ does not contain any edges and thus no vertex has a strictly decreasing trail of length greater than 0. \qed \end{proof}

\noindent We next prove the lower bound on the length of longest strictly decreasing trails.

\begin{theorem}Let $G = (V,E)$ be an undirected edge-weighted graph such that $|V|=n$ and $|E| = q$. Let  
$w:E\rightarrow \{1,\ldots,q\}$ be a weight function assuming different weights are mapped to to different edges. 
Then, $P_d(w,G) \ge 2\cdot\lfloor\frac{q}{n}\rfloor$ i.e., there 
exists a strictly decreasing trail of length $2\cdot\lfloor\frac{q}{n}\rfloor$.\label{theorem:main}\end{theorem}

\begin{proof}Assume that no vertex is a starting point of a trail of length at least $2\cdot\lfloor\frac{q}{n}\rfloor$, that is
	$L^{q}(v) < 2\cdot\lfloor\frac{q}{n}\rfloor,$ for all $v \in V$. 
	Then, $\sum_{v\in V} L^{q}(v) < 2\cdot\lfloor\frac{q}{n}\rfloor n \le 2\cdot q$. But this is a contradiction 
	to Lemma 2 that postulates that the sum of the length of all longest strictly decreasing trails $\sum_{v\in V} L^{q}(v)$ is greater than $2\cdot q$.
	Hence, there has to be at least one vertex with a strictly decreasing trail that is longer than $2\cdot\lfloor\frac{q}{n}\rfloor$ in $G^q$.
	This trail contains a subtrail of length $2\cdot\lfloor\frac{q}{n}\rfloor$. Since $E^q=E$ it follows that $G^q=G$, which concludes 
	the proof. \qed
\end{proof} 

\noindent Based on Theorem \ref{theorem:main}, we get the following results.

\begin{corollary}\label{corollary:IncreasingIsDecreasing}
It holds that $P_i(w,G) \ge 2\cdot\lfloor\frac{q}{n}\rfloor$ since when reversing a strictly decreasing trail 
one obtains a strictly increasing one. In this case, define $L^i(v)$ as the 
length of a longest strictly increasing trail ending at $v$ in $G^i$.\qed \end{corollary}

\begin{corollary}
Let G be as in Theorem \ref{theorem:main} and additionally assume that G is complete. Then, there exists a trail 
of length at least $n-1$, i.e., $f_i(n) = f_d(n) \ge n-1$.\qed
\end{corollary}

In \cite{graham1973increasing} the authors present a non-constructive proof. As in Lemma \ref{lemma:sum} they argue that the 
sum of the lengths of all increasing trails is at least 2. Thus, they overestimate the increase. We however, use the exact increase therefore 
making the proof constructive and obtaining Algorithm \ref{algo:FindLongestTrail}.%
\end{isamarkuptext}\isamarkuptrue%
\isadelimdocument
\endisadelimdocument
\isatagdocument
\isamarkupsection{Formalization of Trail Properties in Isabelle/HOL%
}
\isamarkuptrue%
\endisatagdocument
{\isafolddocument}%
\isadelimdocument
\endisadelimdocument
\begin{isamarkuptext}%
\label{section:Formalization}%
\end{isamarkuptext}\isamarkuptrue%
\isadelimdocument
\endisadelimdocument
\isatagdocument
\isamarkupsubsection{Graph Theory in the Archive of Formal Proofs%
}
\isamarkuptrue%
\endisatagdocument
{\isafolddocument}%
\isadelimdocument
\endisadelimdocument
\begin{isamarkuptext}%
\label{section:GraphTheory} To increase the reusability of our library we build upon the \isa{Graph{\isacharunderscore}Theory}
library by Noschinski \cite{Graph_Theory-AFP}. Graphs are represented as records consisting of vertices and edges that
can be accessed using the selectors \isa{pverts} and \isa{parcs}. We recall the definition 
of the type \isa{pair{\isacharunderscore}pre{\isacharunderscore}digraph}:

\begin{isabelle}%
record\ {\isacharprime}a\ pair{\isacharunderscore}pre{\isacharunderscore}digraph\ {\isacharequal}\ pverts\ {\isacharcolon}{\isacharcolon}\ {\isachardoublequote}{\isacharprime}a\ set{\isachardoublequote}\ parcs\ {\isacharcolon}{\isacharcolon}\ {\isachardoublequote}{\isacharprime}a\ rel{\isachardoublequote}%
\end{isabelle}

Now restrictions upon the two sets and new features can be introduced using locales. 
Locales are Isabelle's way to deal with parameterized theories~\cite{ballarin2010tutorial}. Consider
for example \isa{pair{\isacharunderscore}wf{\isacharunderscore}digraph}.
The endpoints of an edge can be accessed using the functions \isa{fst} and \isa{snd}. Therefore, conditions
\isa{arc{\isacharunderscore}fst{\isacharunderscore}in{\isacharunderscore}verts} and \isa{arc{\isacharunderscore}snd{\isacharunderscore}in{\isacharunderscore}verts} assert that both endpoints of an edge are
vertices. Using so-called sublocales a variety of other graphs are defined.  

\begin{isabelle}%
locale\ pair{\isacharunderscore}wf{\isacharunderscore}digraph\ {\isacharequal}\ pair{\isacharunderscore}pre{\isacharunderscore}digraph\ {\isacharplus}\isanewline
\ \ assumes\ arc{\isacharunderscore}fst{\isacharunderscore}in{\isacharunderscore}verts{\isacharcolon}\ {\isachardoublequote}{\isasymAnd}e{\isachardot}\ e\ {\isasymin}\ parcs\ G\ {\isasymLongrightarrow}\ fst\ e\ {\isasymin}\ pverts\ G{\isachardoublequote}\isanewline
\ \ assumes\ arc{\isacharunderscore}snd{\isacharunderscore}in{\isacharunderscore}verts{\isacharcolon}\ {\isachardoublequote}{\isasymAnd}e{\isachardot}\ e\ {\isasymin}\ parcs\ G\ {\isasymLongrightarrow}\ snd\ e\ {\isasymin}\ pverts\ G{\isachardoublequote}%
\end{isabelle}

An object of type \isa{{\isacharprime}b\ awalk} is defined in \isa{Graph{\isacharunderscore}Theory{\isachardot}Arc{\isacharunderscore}Walk} as a list of edges. 
Additionally, the definition \isa{awalk} imposes that both endpoints of a walk are vertices of 
the graph, all elements of the walk are edges and two subsequent edges share a common vertex. \vspace{1em}

\noindent\isa{type{\isacharunderscore}synonym\ {\isacharprime}b\ awalk\ {\isacharequal}\ {\isachardoublequote}{\isacharprime}b\ list{\isachardoublequote}}

\begin{isabelle}%
definition\ awalk\ {\isacharcolon}{\isacharcolon}\ {\isachardoublequote}{\isacharprime}a\ {\isasymRightarrow}\ {\isacharprime}b\ awalk\ {\isasymRightarrow}\ {\isacharprime}a\ {\isasymRightarrow}\ bool{\isachardoublequote}\isanewline
{\isachardoublequote}awalk\ u\ p\ v\ {\isasymequiv}\ u\ {\isasymin}\ verts\ G\ {\isasymand}\ set\ p\ {\isasymsubseteq}\ arcs\ G\ {\isasymand}\ cas\ u\ p\ v{\isachardoublequote}\ %
\end{isabelle}

\noindent We also reuse the type synonym \isa{weight{\isacharunderscore}fun} introduced in \mbox{\isa{Weighted{\isacharunderscore}Graph}}. \vspace{1em}

\isa{type{\isacharunderscore}synonym\ {\isacharprime}b\ weight{\isacharunderscore}fun\ {\isacharequal}\ {\isachardoublequote}{\isacharprime}b\ {\isasymRightarrow}\ real{\isachardoublequote}}  \vspace{1em}

Finally, there is an useful definition capturing the notion of a complete graph, namely \isa{complete{\isacharunderscore}digraph}.%
\end{isamarkuptext}\isamarkuptrue%
\isadelimdocument
\endisadelimdocument
\isatagdocument
\isamarkupsubsection{Increasing and Decreasing Trails in Weighted Graphs%
}
\isamarkuptrue%
\endisatagdocument
{\isafolddocument}%
\isadelimdocument
\endisadelimdocument
\begin{isamarkuptext}%
\label{section:trails} In our work we extend the graph theory framework from Section \ref{section:GraphTheory} 
with new features enabling reasoning about ordered trails. To this end,
 a trail is defined as a list of edges. We will only consider strictly increasing trails 
on graphs without parallel edges. For this we require the graph 
to be of type \isa{pair{\isacharunderscore}pre{\isacharunderscore}digraph}, as introduced in Section \ref{section:GraphTheory}. 

Two different definitions
are given in our formalization. Function \isa{incTrail} can be used without specifying the first and last vertex of the trail
whereas \isa{incTrail{\isadigit{2}}} uses more of \isa{Graph{\isacharunderscore}Theory{\isacharprime}s} predefined features. Moreover, making use of  monotonicity
\mbox{\isa{incTrail}} only requires to check if one edge's weight is smaller than its successors' while \isa{incTrail{\isadigit{2}}}
checks if the weight is smaller than the one of all subsequent edges in the sequence, i.e. if the list 
is sorted. The {\em equivalence
between the two notions} is shown in the following.%
\end{isamarkuptext}\isamarkuptrue%
\isacommand{fun}\isamarkupfalse%
\ incTrail\ {\isacharcolon}{\isacharcolon}\ {\isachardoublequoteopen}{\isacharprime}a\ pair{\isacharunderscore}pre{\isacharunderscore}digraph\ {\isasymRightarrow}\ {\isacharparenleft}{\isacharprime}a\ {\isasymtimes}{\isacharprime}a{\isacharparenright}\ weight{\isacharunderscore}fun\ {\isasymRightarrow}\ {\isacharparenleft}{\isacharprime}a\ {\isasymtimes}{\isacharprime}a{\isacharparenright}\ list\ {\isasymRightarrow}\ bool{\isachardoublequoteclose}\ \isakeyword{where}\isanewline
{\isachardoublequoteopen}incTrail\ g\ w\ {\isacharbrackleft}{\isacharbrackright}\ {\isacharequal}\ True{\isachardoublequoteclose}\ {\isacharbar}\isanewline
{\isachardoublequoteopen}incTrail\ g\ w\ {\isacharbrackleft}e\isactrlsub {\isadigit{1}}{\isacharbrackright}\ {\isacharequal}\ {\isacharparenleft}e\isactrlsub {\isadigit{1}}\ {\isasymin}\ parcs\ g{\isacharparenright}{\isachardoublequoteclose}\ {\isacharbar}\isanewline
{\isachardoublequoteopen}incTrail\ g\ w\ {\isacharparenleft}e\isactrlsub {\isadigit{1}}{\isacharhash}e\isactrlsub {\isadigit{2}}{\isacharhash}es{\isacharparenright}\ {\isacharequal}\ {\isacharparenleft}if\ w\ e\isactrlsub {\isadigit{1}}\ {\isacharless}\ w\ e\isactrlsub {\isadigit{2}}\ {\isasymand}\ e\isactrlsub {\isadigit{1}}\ {\isasymin}\ parcs\ g\ {\isasymand}\ snd\ e\isactrlsub {\isadigit{1}}\ {\isacharequal}\ fst\ e\isactrlsub {\isadigit{2}}\ \isanewline
\ \ \ \ \ \ \ \ \ \ \ \ \ \ \ \ \ \ \ \ \ \ \ \ \ \ \ \ \ \ \ \ \ \ \ \ then\ incTrail\ g\ w\ {\isacharparenleft}e\isactrlsub {\isadigit{2}}{\isacharhash}es{\isacharparenright}\ else\ False{\isacharparenright}{\isachardoublequoteclose}\isanewline
\isanewline
\isacommand{definition}\isamarkupfalse%
{\isacharparenleft}\isakeyword{in}\ pair{\isacharunderscore}pre{\isacharunderscore}digraph{\isacharparenright}\ incTrail{\isadigit{2}}\ \isakeyword{where}\isanewline
{\isachardoublequoteopen}incTrail{\isadigit{2}}\ w\ es\ u\ v\ {\isasymequiv}\ sorted{\isacharunderscore}wrt\ {\isacharparenleft}{\isasymlambda}\ e\isactrlsub {\isadigit{1}}\ e\isactrlsub {\isadigit{2}}{\isachardot}\ w\ e\isactrlsub {\isadigit{1}}\ {\isacharless}\ w\ e\isactrlsub {\isadigit{2}}{\isacharparenright}\ es\ {\isasymand}\ {\isacharparenleft}es\ {\isacharequal}\ {\isacharbrackleft}{\isacharbrackright}\ {\isasymor}\ awalk\ u\ es\ v{\isacharparenright}{\isachardoublequoteclose}\isanewline
\isanewline
\isacommand{fun}\isamarkupfalse%
\ decTrail\ {\isacharcolon}{\isacharcolon}\ {\isachardoublequoteopen}{\isacharprime}a\ pair{\isacharunderscore}pre{\isacharunderscore}digraph\ {\isasymRightarrow}\ {\isacharparenleft}{\isacharprime}a\ {\isasymtimes}{\isacharprime}a{\isacharparenright}\ weight{\isacharunderscore}fun\ {\isasymRightarrow}\ {\isacharparenleft}{\isacharprime}a\ {\isasymtimes}{\isacharprime}a{\isacharparenright}\ list\ {\isasymRightarrow}\ bool{\isachardoublequoteclose}\ \isakeyword{where}\isanewline
{\isachardoublequoteopen}decTrail\ g\ w\ {\isacharbrackleft}{\isacharbrackright}\ {\isacharequal}\ True{\isachardoublequoteclose}\ {\isacharbar}\isanewline
{\isachardoublequoteopen}decTrail\ g\ w\ {\isacharbrackleft}e\isactrlsub {\isadigit{1}}{\isacharbrackright}\ {\isacharequal}\ {\isacharparenleft}e\isactrlsub {\isadigit{1}}\ {\isasymin}\ parcs\ g{\isacharparenright}{\isachardoublequoteclose}\ {\isacharbar}\isanewline
{\isachardoublequoteopen}decTrail\ g\ w\ {\isacharparenleft}e\isactrlsub {\isadigit{1}}{\isacharhash}e\isactrlsub {\isadigit{2}}{\isacharhash}es{\isacharparenright}\ {\isacharequal}\ {\isacharparenleft}if\ w\ e\isactrlsub {\isadigit{1}}\ {\isachargreater}\ w\ e\isactrlsub {\isadigit{2}}\ {\isasymand}\ e\isactrlsub {\isadigit{1}}\ {\isasymin}\ parcs\ g\ {\isasymand}\ snd\ e\isactrlsub {\isadigit{1}}\ {\isacharequal}\ fst\ e\isactrlsub {\isadigit{2}}\ \isanewline
\ \ \ \ \ \ \ \ \ \ \ \ \ \ \ \ \ \ \ \ \ \ \ \ \ \ \ \ \ \ \ \ \ \ \ \ then\ decTrail\ g\ w\ {\isacharparenleft}e\isactrlsub {\isadigit{2}}{\isacharhash}es{\isacharparenright}\ else\ False{\isacharparenright}{\isachardoublequoteclose}\isanewline
\isanewline
\isacommand{definition}\isamarkupfalse%
{\isacharparenleft}\isakeyword{in}\ pair{\isacharunderscore}pre{\isacharunderscore}digraph{\isacharparenright}\ decTrail{\isadigit{2}}\ \isakeyword{where}\ \isanewline
{\isachardoublequoteopen}decTrail{\isadigit{2}}\ w\ es\ u\ v\ {\isasymequiv}\ sorted{\isacharunderscore}wrt\ {\isacharparenleft}{\isasymlambda}\ e\isactrlsub {\isadigit{1}}\ e\isactrlsub {\isadigit{2}}{\isachardot}\ w\ e\isactrlsub {\isadigit{1}}\ {\isachargreater}\ w\ e\isactrlsub {\isadigit{2}}{\isacharparenright}\ es\ {\isasymand}\ {\isacharparenleft}es\ {\isacharequal}\ {\isacharbrackleft}{\isacharbrackright}\ {\isasymor}\ awalk\ u\ es\ v{\isacharparenright}{\isachardoublequoteclose}%
\begin{isamarkuptext}%
Defining trails as lists in Isabelle has many advantages including using predefined list operators, 
e.g., drop. Thus, we can show one result that will be constantly needed in the following, that is, that
{\em any subtrail of an ordered trail is an ordered trail itself}.%
\end{isamarkuptext}\isamarkuptrue%
\isacommand{lemma}\isamarkupfalse%
\ incTrail{\isacharunderscore}subtrail{\isacharcolon}\isanewline
\ \ \isakeyword{assumes}\ {\isachardoublequoteopen}incTrail\ g\ w\ es{\isachardoublequoteclose}\isanewline
\ \ \isakeyword{shows}\ {\isachardoublequoteopen}incTrail\ g\ w\ {\isacharparenleft}drop\ k\ es{\isacharparenright}{\isachardoublequoteclose}%
\isadelimproof
\endisadelimproof
\isatagproof
\endisatagproof
{\isafoldproof}%
\isadelimproof
\endisadelimproof
\begin{isamarkuptext}%
\end{isamarkuptext}\isamarkuptrue%
\isacommand{lemma}\isamarkupfalse%
\ decTrail{\isacharunderscore}subtrail{\isacharcolon}\ \isanewline
\ \ \isakeyword{assumes}\ {\isachardoublequoteopen}decTrail\ g\ w\ es{\isachardoublequoteclose}\isanewline
\ \ \isakeyword{shows}\ {\isachardoublequoteopen}decTrail\ g\ w\ {\isacharparenleft}drop\ k\ es{\isacharparenright}{\isachardoublequoteclose}%
\isadelimproof
\endisadelimproof
\isatagproof
\endisatagproof
{\isafoldproof}%
\isadelimproof
\endisadelimproof
\isadelimproof
\endisadelimproof
\isatagproof
\endisatagproof
{\isafoldproof}%
\isadelimproof
\endisadelimproof
\isadelimproof
\endisadelimproof
\isatagproof
\endisatagproof
{\isafoldproof}%
\isadelimproof
\endisadelimproof
\isadelimproof
\endisadelimproof
\isatagproof
\endisatagproof
{\isafoldproof}%
\isadelimproof
\endisadelimproof
\isadelimproof
\endisadelimproof
\isatagproof
\endisatagproof
{\isafoldproof}%
\isadelimproof
\endisadelimproof
\isadelimproof
\endisadelimproof
\isatagproof
\endisatagproof
{\isafoldproof}%
\isadelimproof
\endisadelimproof
\isadelimproof
\endisadelimproof
\isatagproof
\endisatagproof
{\isafoldproof}%
\isadelimproof
\endisadelimproof
\isadelimproof
\endisadelimproof
\isatagproof
\endisatagproof
{\isafoldproof}%
\isadelimproof
\endisadelimproof
\isadelimproof
\endisadelimproof
\isatagproof
\endisatagproof
{\isafoldproof}%
\isadelimproof
\endisadelimproof
\isadelimproof
\endisadelimproof
\isatagproof
\endisatagproof
{\isafoldproof}%
\isadelimproof
\endisadelimproof
\isadelimproof
\endisadelimproof
\isatagproof
\endisatagproof
{\isafoldproof}%
\isadelimproof
\endisadelimproof
\begin{isamarkuptext}%
In Isabelle we then show the equivalence between the two definitions \isa{decTrail} and \isa{decTrail{\isadigit{2}}} of strictly decreasing trails.
Similarly, we also show the equivalence between the definition \isa{incTrail} and \isa{incTrail{\isadigit{2}}} of strictly increasing trails.%
\end{isamarkuptext}\isamarkuptrue%
\isacommand{lemma}\isamarkupfalse%
{\isacharparenleft}\isakeyword{in}\ pair{\isacharunderscore}wf{\isacharunderscore}digraph{\isacharparenright}\ decTrail{\isacharunderscore}is{\isacharunderscore}dec{\isacharunderscore}walk{\isacharcolon}\isanewline
\ \ \isakeyword{shows}\ {\isachardoublequoteopen}decTrail\ G\ w\ es\ {\isasymlongleftrightarrow}\ decTrail{\isadigit{2}}\ w\ es\ {\isacharparenleft}fst\ {\isacharparenleft}hd\ es{\isacharparenright}{\isacharparenright}\ {\isacharparenleft}snd\ {\isacharparenleft}last\ es{\isacharparenright}{\isacharparenright}{\isachardoublequoteclose}%
\isadelimproof
\endisadelimproof
\isatagproof
\endisatagproof
{\isafoldproof}%
\isadelimproof
\endisadelimproof
\begin{isamarkuptext}%
\end{isamarkuptext}\isamarkuptrue%
\isacommand{lemma}\isamarkupfalse%
{\isacharparenleft}\isakeyword{in}\ pair{\isacharunderscore}wf{\isacharunderscore}digraph{\isacharparenright}\ incTrail{\isacharunderscore}is{\isacharunderscore}inc{\isacharunderscore}walk{\isacharcolon}\isanewline
\ \ \isakeyword{shows}\ {\isachardoublequoteopen}incTrail\ G\ w\ es\ {\isasymlongleftrightarrow}\ incTrail{\isadigit{2}}\ w\ es\ {\isacharparenleft}fst\ {\isacharparenleft}hd\ es{\isacharparenright}{\isacharparenright}\ {\isacharparenleft}snd\ {\isacharparenleft}last\ es{\isacharparenright}{\isacharparenright}{\isachardoublequoteclose}%
\isadelimproof
\endisadelimproof
\isatagproof
\endisatagproof
{\isafoldproof}%
\isadelimproof
\endisadelimproof
\isadelimproof
\endisadelimproof
\isatagproof
\endisatagproof
{\isafoldproof}%
\isadelimproof
\endisadelimproof
\isadelimproof
\endisadelimproof
\isatagproof
\endisatagproof
{\isafoldproof}%
\isadelimproof
\endisadelimproof
\begin{isamarkuptext}%
Any strictly decreasing trail $(e_1,\ldots,e_n)$ can also be seen as a strictly increasing trail $(e_n,...,e_1)$
if the graph considered is undirected. To this end, we make use of the locale \isa{pair{\isacharunderscore}sym{\isacharunderscore}digraph}
that captures the idea of symmetric arcs. However, it is also necessary to assume that the weight 
function assigns the same weight to edge $(v_i,v_j)$ as to $(v_j,v_i)$. This assumption is therefore
added to \isa{decTrail{\isacharunderscore}eq{\isacharunderscore}rev{\isacharunderscore}incTrail} and \isa{incTrail{\isacharunderscore}eq{\isacharunderscore}rev{\isacharunderscore}decTrail}.%
\end{isamarkuptext}\isamarkuptrue%
\isacommand{lemma}\isamarkupfalse%
{\isacharparenleft}\isakeyword{in}\ pair{\isacharunderscore}sym{\isacharunderscore}digraph{\isacharparenright}\ decTrail{\isacharunderscore}eq{\isacharunderscore}rev{\isacharunderscore}incTrail{\isacharcolon}\isanewline
\ \ \isakeyword{assumes}\ {\isachardoublequoteopen}{\isasymforall}\ v\isactrlsub {\isadigit{1}}\ v\isactrlsub {\isadigit{2}}{\isachardot}\ w\ {\isacharparenleft}v\isactrlsub {\isadigit{1}}{\isacharcomma}v\isactrlsub {\isadigit{2}}{\isacharparenright}\ {\isacharequal}\ w{\isacharparenleft}v\isactrlsub {\isadigit{2}}{\isacharcomma}v\isactrlsub {\isadigit{1}}{\isacharparenright}{\isachardoublequoteclose}\ \isanewline
\ \ \isakeyword{shows}\ {\isachardoublequoteopen}decTrail\ G\ w\ es\ {\isasymlongleftrightarrow}\ incTrail\ G\ w\ {\isacharparenleft}rev\ {\isacharparenleft}map\ {\isacharparenleft}{\isasymlambda}{\isacharparenleft}v\isactrlsub {\isadigit{1}}{\isacharcomma}v\isactrlsub {\isadigit{2}}{\isacharparenright}{\isachardot}\ {\isacharparenleft}v\isactrlsub {\isadigit{2}}{\isacharcomma}v\isactrlsub {\isadigit{1}}{\isacharparenright}{\isacharparenright}\ es{\isacharparenright}{\isacharparenright}{\isachardoublequoteclose}\isanewline
\ \ \ \ %
\isadelimproof
\endisadelimproof
\isatagproof
\endisatagproof
{\isafoldproof}%
\isadelimproof
\endisadelimproof
\isanewline
\isacommand{lemma}\isamarkupfalse%
{\isacharparenleft}\isakeyword{in}\ pair{\isacharunderscore}sym{\isacharunderscore}digraph{\isacharparenright}\ incTrail{\isacharunderscore}eq{\isacharunderscore}rev{\isacharunderscore}decTrail{\isacharcolon}\isanewline
\ \ \isakeyword{assumes}\ {\isachardoublequoteopen}{\isasymforall}\ v\isactrlsub {\isadigit{1}}\ v\isactrlsub {\isadigit{2}}{\isachardot}\ w\ {\isacharparenleft}v\isactrlsub {\isadigit{1}}{\isacharcomma}v\isactrlsub {\isadigit{2}}{\isacharparenright}\ {\isacharequal}\ w{\isacharparenleft}v\isactrlsub {\isadigit{2}}{\isacharcomma}v\isactrlsub {\isadigit{1}}{\isacharparenright}{\isachardoublequoteclose}\ \isanewline
\ \ \isakeyword{shows}\ {\isachardoublequoteopen}incTrail\ G\ w\ es\ {\isasymlongleftrightarrow}\ decTrail\ G\ w\ {\isacharparenleft}rev\ {\isacharparenleft}map\ {\isacharparenleft}{\isasymlambda}{\isacharparenleft}v\isactrlsub {\isadigit{1}}{\isacharcomma}v\isactrlsub {\isadigit{2}}{\isacharparenright}{\isachardot}\ {\isacharparenleft}v\isactrlsub {\isadigit{2}}{\isacharcomma}v\isactrlsub {\isadigit{1}}{\isacharparenright}{\isacharparenright}\ es{\isacharparenright}{\isacharparenright}{\isachardoublequoteclose}%
\isadelimproof
\endisadelimproof
\isatagproof
\endisatagproof
{\isafoldproof}%
\isadelimproof
\endisadelimproof
\isadelimdocument
\endisadelimdocument
\isatagdocument
\isamarkupsubsection{Weighted Graphs%
}
\isamarkuptrue%
\endisatagdocument
{\isafolddocument}%
\isadelimdocument
\endisadelimdocument
\begin{isamarkuptext}%
\label{section:localeSurjective} We add the locale \isa{weighted{\isacharunderscore}pair{\isacharunderscore}graph} 
on top of the locale \isa{pair{\isacharunderscore}graph} introduced in \isa{Graph{\isacharunderscore}Theory}. A  \isa{pair{\isacharunderscore}graph} is a 
finite, loop free and symmetric graph. We do not restrict the types of vertices and edges but impose 
the condition that they have to be a linear order.

 Furthermore, all weights have to be integers between 0 and $\lfloor\frac{q}{2}\rfloor$ where 0 is 
used as a special value to indicate that there is no edge at that position. Since the 
range of the weight function is in the reals, the set of natural numbers
\mbox{\isa{{\isacharbraceleft}{\isadigit{1}}{\isacharcomma}{\isachardot}{\isachardot}{\isacharcomma}card\ {\isacharparenleft}parcs\ G{\isacharparenright}\ div\ {\isadigit{2}}{\isacharbraceright}}} has to be casted into a set of reals. This is realized by taking the image
of the function \isa{real} that casts a natural number to a real.%
\end{isamarkuptext}\isamarkuptrue%
\isacommand{locale}\isamarkupfalse%
\ weighted{\isacharunderscore}pair{\isacharunderscore}graph\ {\isacharequal}\ pair{\isacharunderscore}graph\ {\isachardoublequoteopen}{\isacharparenleft}G{\isacharcolon}{\isacharcolon}\ {\isacharparenleft}{\isacharprime}a{\isacharcolon}{\isacharcolon}linorder{\isacharparenright}\ pair{\isacharunderscore}pre{\isacharunderscore}digraph{\isacharparenright}{\isachardoublequoteclose}\ \isakeyword{for}\ G\ {\isacharplus}\isanewline
\ \ \isakeyword{fixes}\ w\ {\isacharcolon}{\isacharcolon}\ {\isachardoublequoteopen}{\isacharparenleft}{\isacharprime}a{\isasymtimes}{\isacharprime}a{\isacharparenright}\ weight{\isacharunderscore}fun{\isachardoublequoteclose}\isanewline
\ \ \isakeyword{assumes}\ dom{\isacharcolon}\ {\isachardoublequoteopen}e\ {\isasymin}\ parcs\ G\ {\isasymlongrightarrow}\ w\ e\ {\isasymin}\ real\ {\isacharbackquote}\ {\isacharbraceleft}{\isadigit{1}}{\isachardot}{\isachardot}card\ {\isacharparenleft}parcs\ G{\isacharparenright}\ div\ {\isadigit{2}}{\isacharbraceright}{\isachardoublequoteclose}\ \isanewline
\ \ \ \ \ \ \isakeyword{and}\ vert{\isacharunderscore}ge{\isacharcolon}\ {\isachardoublequoteopen}card\ {\isacharparenleft}pverts\ G{\isacharparenright}\ {\isasymge}\ {\isadigit{1}}{\isachardoublequoteclose}%
\begin{isamarkuptext}%
We introduce some useful abbreviations, according to the ones in Section \ref{section:Prelim}%
\end{isamarkuptext}\isamarkuptrue%
\isacommand{abbreviation}\isamarkupfalse%
{\isacharparenleft}\isakeyword{in}\ weighted{\isacharunderscore}pair{\isacharunderscore}graph{\isacharparenright}\ {\isachardoublequoteopen}q\ {\isasymequiv}\ card\ {\isacharparenleft}parcs\ G{\isacharparenright}{\isachardoublequoteclose}\isanewline
\isacommand{abbreviation}\isamarkupfalse%
{\isacharparenleft}\isakeyword{in}\ weighted{\isacharunderscore}pair{\isacharunderscore}graph{\isacharparenright}\ {\isachardoublequoteopen}n\ {\isasymequiv}\ card\ {\isacharparenleft}pverts\ G{\isacharparenright}{\isachardoublequoteclose}\isanewline
\isacommand{abbreviation}\isamarkupfalse%
{\isacharparenleft}\isakeyword{in}\ weighted{\isacharunderscore}pair{\isacharunderscore}graph{\isacharparenright}\ {\isachardoublequoteopen}W\ {\isasymequiv}\ {\isacharbraceleft}{\isadigit{1}}{\isachardot}{\isachardot}q\ div\ {\isadigit{2}}{\isacharbraceright}{\isachardoublequoteclose}%
\begin{isamarkuptext}%
Note an important difference between Section \ref{section:symbolicProof} and our formalization. Although 
a \isa{weighted{\isacharunderscore}pair{\isacharunderscore}graph} is symmetric, the edge set contains both ``directions" of an edge, 
i.e., $(v_1,v_2)$ and $(v_2,v_1)$ are both in \isa{parcs\ G}. Thus, the maximum number of edges (in the 
case that the graph is complete) is $n\cdot(n-1)$ and not $\frac{n\cdot(n-1)}{2}$. Another consequence is that
the number $q$ of edges is always even.%
\end{isamarkuptext}\isamarkuptrue%
\isacommand{lemma}\isamarkupfalse%
\ {\isacharparenleft}\isakeyword{in}\ weighted{\isacharunderscore}pair{\isacharunderscore}graph{\isacharparenright}\ max{\isacharunderscore}arcs{\isacharcolon}\ \isanewline
\ \ \isakeyword{shows}\ {\isachardoublequoteopen}card\ {\isacharparenleft}parcs\ G{\isacharparenright}\ {\isasymle}\ n{\isacharasterisk}{\isacharparenleft}n{\isacharminus}{\isadigit{1}}{\isacharparenright}{\isachardoublequoteclose}%
\isadelimproof
\endisadelimproof
\isatagproof
\endisatagproof
{\isafoldproof}%
\isadelimproof
\endisadelimproof
\isadelimproof
\endisadelimproof
\isatagproof
\endisatagproof
{\isafoldproof}%
\isadelimproof
\endisadelimproof
\isadelimproof
\endisadelimproof
\isatagproof
\endisatagproof
{\isafoldproof}%
\isadelimproof
\endisadelimproof
\isadelimproof
\endisadelimproof
\isatagproof
\endisatagproof
{\isafoldproof}%
\isadelimproof
\endisadelimproof
\begin{isamarkuptext}%
\end{isamarkuptext}\isamarkuptrue%
\isacommand{lemma}\isamarkupfalse%
\ {\isacharparenleft}\isakeyword{in}\ weighted{\isacharunderscore}pair{\isacharunderscore}graph{\isacharparenright}\ even{\isacharunderscore}arcs{\isacharcolon}\ \isanewline
\ \ \isakeyword{shows}\ {\isachardoublequoteopen}even\ q{\isachardoublequoteclose}%
\isadelimproof
\endisadelimproof
\isatagproof
\endisatagproof
{\isafoldproof}%
\isadelimproof
\endisadelimproof
\begin{isamarkuptext}%
The below sublocale \isa{distinct{\isacharunderscore}weighted{\isacharunderscore}pair{\isacharunderscore}graph} refines
\isa{weighted{\isacharunderscore}pair{\isacharunderscore}graph}. The condition 
\isa{zero} fixes the meaning of 0.
The weight function is defined on the set of all vertices but since self loops are not allowed; 
we use 0 as a special value to indicate the unavailability of the edge. 
The second condition \isa{distinct} enforces that no two edges can have the same weight. There
are some exceptions however captured in the statement \isa{{\isacharparenleft}v\isactrlsub {\isadigit{1}}\ {\isacharequal}\ u\isactrlsub {\isadigit{2}}\ {\isasymand}\ v\isactrlsub {\isadigit{2}}\ {\isacharequal}\ u\isactrlsub {\isadigit{1}}{\isacharparenright}\ {\isasymor}\ {\isacharparenleft}v\isactrlsub {\isadigit{1}}\ {\isacharequal}\ u\isactrlsub {\isadigit{1}}\ {\isasymand}\ v\isactrlsub {\isadigit{2}}\ {\isacharequal}\ u\isactrlsub {\isadigit{2}}{\isacharparenright}}.
Firstly, $(v_1,v_2)$ should have the same weight as $(v_2,v_1)$. Secondly, $w(v_1,v_2)$ has the
same value as $w(v_1,v_2)$. Note that both edges being self loops resulting in them both having 
weight 0 is prohibited by condition \isa{zero}.
Our decision to separate these two conditions from the ones in \isa{weighted{\isacharunderscore}pair{\isacharunderscore}graph}
instead of making one locale of its own is two-fold: On the one hand, there are scenarios where 
distinctiveness is not wished for. On the other hand, 0 might not be available as a special value.%
\end{isamarkuptext}\isamarkuptrue%
\isacommand{locale}\isamarkupfalse%
\ distinct{\isacharunderscore}weighted{\isacharunderscore}pair{\isacharunderscore}graph\ {\isacharequal}\ weighted{\isacharunderscore}pair{\isacharunderscore}graph\ {\isacharplus}\ \isanewline
\ \ \isakeyword{assumes}\ zero{\isacharcolon}\ {\isachardoublequoteopen}{\isasymforall}\ v\isactrlsub {\isadigit{1}}\ v\isactrlsub {\isadigit{2}}{\isachardot}\ {\isacharparenleft}v\isactrlsub {\isadigit{1}}{\isacharcomma}v\isactrlsub {\isadigit{2}}{\isacharparenright}\ {\isasymnotin}\ parcs\ G\ {\isasymlongleftrightarrow}\ w\ {\isacharparenleft}v\isactrlsub {\isadigit{1}}{\isacharcomma}v\isactrlsub {\isadigit{2}}{\isacharparenright}\ {\isacharequal}\ {\isadigit{0}}{\isachardoublequoteclose}\isanewline
\ \ \ \ \ \ \isakeyword{and}\ distinct{\isacharcolon}\ {\isachardoublequoteopen}{\isasymforall}\ {\isacharparenleft}v\isactrlsub {\isadigit{1}}{\isacharcomma}v\isactrlsub {\isadigit{2}}{\isacharparenright}\ {\isasymin}\ parcs\ G{\isachardot}\ {\isasymforall}\ {\isacharparenleft}u\isactrlsub {\isadigit{1}}{\isacharcomma}u\isactrlsub {\isadigit{2}}{\isacharparenright}\ {\isasymin}\ parcs\ G{\isachardot}\ \isanewline
\ \ \ \ \ \ {\isacharparenleft}{\isacharparenleft}v\isactrlsub {\isadigit{1}}\ {\isacharequal}\ u\isactrlsub {\isadigit{2}}\ {\isasymand}\ v\isactrlsub {\isadigit{2}}\ {\isacharequal}\ u\isactrlsub {\isadigit{1}}{\isacharparenright}\ {\isasymor}\ {\isacharparenleft}v\isactrlsub {\isadigit{1}}\ {\isacharequal}\ u\isactrlsub {\isadigit{1}}\ {\isasymand}\ v\isactrlsub {\isadigit{2}}\ {\isacharequal}\ u\isactrlsub {\isadigit{2}}{\isacharparenright}{\isacharparenright}\ {\isasymlongleftrightarrow}\ w\ {\isacharparenleft}v\isactrlsub {\isadigit{1}}{\isacharcomma}v\isactrlsub {\isadigit{2}}{\isacharparenright}\ {\isacharequal}\ w\ {\isacharparenleft}u\isactrlsub {\isadigit{1}}{\isacharcomma}u\isactrlsub {\isadigit{2}}{\isacharparenright}{\isachardoublequoteclose}%
\isadelimproof
\endisadelimproof
\isatagproof
\endisatagproof
{\isafoldproof}%
\isadelimproof
\endisadelimproof
\isadelimproof
\endisadelimproof
\isatagproof
\endisatagproof
{\isafoldproof}%
\isadelimproof
\endisadelimproof
\isadelimproof
\endisadelimproof
\isatagproof
\endisatagproof
{\isafoldproof}%
\isadelimproof
\endisadelimproof
\isadelimproof
\endisadelimproof
\isatagproof
\endisatagproof
{\isafoldproof}%
\isadelimproof
\endisadelimproof
\isadelimproof
\endisadelimproof
\isatagproof
\endisatagproof
{\isafoldproof}%
\isadelimproof
\endisadelimproof
\isadelimproof
\endisadelimproof
\isatagproof
\endisatagproof
{\isafoldproof}%
\isadelimproof
\endisadelimproof
\isadelimproof
\endisadelimproof
\isatagproof
\endisatagproof
{\isafoldproof}%
\isadelimproof
\endisadelimproof
\isadelimproof
\endisadelimproof
\isatagproof
\endisatagproof
{\isafoldproof}%
\isadelimproof
\endisadelimproof
\begin{isamarkuptext}%
One important step in our formalization is to show that the weight function is surjective. However, having two 
elements of the domain (edges) being mapped to the same element of the codomain (weight) makes 
the proof complicated. We therefore first prove that the weight function is surjective on a restricted
set of edges. Here we use the fact that there is a linear order on vertices by only considering edges
were the first endpoint is bigger than the second. 

Then, the surjectivity of $w$ is relatively simple to show. Note that we could also have assumed surjectivity in 
\isa{distinct{\isacharunderscore}weighted{\isacharunderscore}pair{\isacharunderscore}graph} and shown that distinctiveness follows from it. However,
distinctiveness is the more natural assumption that is more likely to appear in any application
of ordered trails.%
\end{isamarkuptext}\isamarkuptrue%
\isacommand{lemma}\isamarkupfalse%
{\isacharparenleft}\isakeyword{in}\ distinct{\isacharunderscore}weighted{\isacharunderscore}pair{\isacharunderscore}graph{\isacharparenright}\ restricted{\isacharunderscore}weight{\isacharunderscore}fun{\isacharunderscore}surjective{\isacharcolon}\ \ \isanewline
\ \ {\isachardoublequoteopen}{\isasymforall}k\ {\isasymin}\ W{\isachardot}\ {\isasymexists}{\isacharparenleft}v\isactrlsub {\isadigit{1}}{\isacharcomma}v\isactrlsub {\isadigit{2}}{\isacharparenright}\ {\isasymin}\ {\isacharbraceleft}{\isacharparenleft}p{\isadigit{1}}{\isacharcomma}p{\isadigit{2}}{\isacharparenright}{\isachardot}\ {\isacharparenleft}p{\isadigit{1}}{\isacharcomma}p{\isadigit{2}}{\isacharparenright}\ {\isasymin}\ parcs\ G\ {\isasymand}\ p{\isadigit{2}}\ {\isacharless}\ p{\isadigit{1}}{\isacharbraceright}{\isachardot}\ w\ {\isacharparenleft}v\isactrlsub {\isadigit{1}}{\isacharcomma}v\isactrlsub {\isadigit{2}}{\isacharparenright}\ {\isacharequal}\ k{\isachardoublequoteclose}%
\isadelimproof
\endisadelimproof
\isatagproof
\endisatagproof
{\isafoldproof}%
\isadelimproof
\endisadelimproof
\begin{isamarkuptext}%
\end{isamarkuptext}\isamarkuptrue%
\isacommand{lemma}\isamarkupfalse%
{\isacharparenleft}\isakeyword{in}\ distinct{\isacharunderscore}weighted{\isacharunderscore}pair{\isacharunderscore}graph{\isacharparenright}\ weight{\isacharunderscore}fun{\isacharunderscore}surjective{\isacharcolon}\isanewline
\ \ \isakeyword{shows}\ {\isachardoublequoteopen}{\isasymforall}k\ {\isasymin}\ W{\isachardot}\ {\isasymexists}{\isacharparenleft}v\isactrlsub {\isadigit{1}}{\isacharcomma}v\isactrlsub {\isadigit{2}}{\isacharparenright}\ {\isasymin}\ parcs\ G{\isachardot}\ w\ {\isacharparenleft}v\isactrlsub {\isadigit{1}}{\isacharcomma}v\isactrlsub {\isadigit{2}}{\isacharparenright}\ {\isacharequal}\ k{\isachardoublequoteclose}%
\isadelimproof
\endisadelimproof
\isatagproof
\endisatagproof
{\isafoldproof}%
\isadelimproof
\endisadelimproof
\isadelimproof
\endisadelimproof
\isatagproof
\endisatagproof
{\isafoldproof}%
\isadelimproof
\endisadelimproof
\isadelimproof
\endisadelimproof
\isatagproof
\endisatagproof
{\isafoldproof}%
\isadelimproof
\endisadelimproof
\isadelimproof
\endisadelimproof
\isatagproof
\endisatagproof
{\isafoldproof}%
\isadelimproof
\endisadelimproof
\isadelimproof
\endisadelimproof
\isatagproof
\endisatagproof
{\isafoldproof}%
\isadelimproof
\endisadelimproof
\isadelimproof
\endisadelimproof
\isatagproof
\endisatagproof
{\isafoldproof}%
\isadelimproof
\endisadelimproof
\isadelimdocument
\endisadelimdocument
\isatagdocument
\isamarkupsubsection{Computing a Longest Ordered Trail%
}
\isamarkuptrue%
\endisatagdocument
{\isafolddocument}%
\isadelimdocument
\endisadelimdocument
\begin{isamarkuptext}%
\label{section:computeLongestTrail}We next formally verify Algorithm \ref{algo:FindLongestTrail} and compute longest ordered trails. To this end, 
we introduce the function \isa{findEdge} to find an edge in a list of edges by its weight.%
\end{isamarkuptext}\isamarkuptrue%
\isacommand{fun}\isamarkupfalse%
\ findEdge\ {\isacharcolon}{\isacharcolon}\ {\isachardoublequoteopen}{\isacharparenleft}{\isacharprime}a{\isasymtimes}{\isacharprime}a{\isacharparenright}\ weight{\isacharunderscore}fun\ {\isasymRightarrow}\ {\isacharparenleft}{\isacharprime}a{\isasymtimes}{\isacharprime}a{\isacharparenright}\ list\ {\isasymRightarrow}\ real\ {\isasymRightarrow}\ {\isacharparenleft}{\isacharprime}a{\isasymtimes}{\isacharprime}a{\isacharparenright}{\isachardoublequoteclose}\ \isakeyword{where}\isanewline
{\isachardoublequoteopen}findEdge\ f\ {\isacharbrackleft}{\isacharbrackright}\ k\ {\isacharequal}\ undefined{\isachardoublequoteclose}\ {\isacharbar}\isanewline
{\isachardoublequoteopen}findEdge\ f\ {\isacharparenleft}e{\isacharhash}es{\isacharparenright}\ k\ {\isacharequal}\ {\isacharparenleft}if\ f\ e\ {\isacharequal}\ k\ then\ e\ else\ findEdge\ f\ es\ k{\isacharparenright}{\isachardoublequoteclose}%
\isadelimproof
\endisadelimproof
\isatagproof
\endisatagproof
{\isafoldproof}%
\isadelimproof
\endisadelimproof
\begin{isamarkuptext}%
Function \isa{findEdge} will correctly return the edge whose weight is $k$. We do not care in which order the endpoints
are found, i.e. whether $(v_1,v_2)$ or $(v_2,v_1)$ is returned.%
\end{isamarkuptext}\isamarkuptrue%
\isacommand{lemma}\isamarkupfalse%
{\isacharparenleft}\isakeyword{in}\ distinct{\isacharunderscore}weighted{\isacharunderscore}pair{\isacharunderscore}graph{\isacharparenright}\ findEdge{\isacharunderscore}success{\isacharcolon}\isanewline
\ \ \isakeyword{assumes}\ {\isachardoublequoteopen}k\ {\isasymin}\ W{\isachardoublequoteclose}\ \isakeyword{and}\ {\isachardoublequoteopen}w\ {\isacharparenleft}v\isactrlsub {\isadigit{1}}{\isacharcomma}v\isactrlsub {\isadigit{2}}{\isacharparenright}\ {\isacharequal}\ k{\isachardoublequoteclose}\ \isakeyword{and}\ {\isachardoublequoteopen}{\isacharparenleft}parcs\ G{\isacharparenright}\ {\isasymnoteq}\ {\isacharbraceleft}{\isacharbraceright}{\isachardoublequoteclose}\ \isanewline
\ \ \isakeyword{shows}\ {\isachardoublequoteopen}{\isacharparenleft}findEdge\ w\ {\isacharparenleft}set{\isacharunderscore}to{\isacharunderscore}list\ {\isacharparenleft}parcs\ G{\isacharparenright}{\isacharparenright}\ k{\isacharparenright}\ {\isacharequal}\ {\isacharparenleft}v\isactrlsub {\isadigit{1}}{\isacharcomma}v\isactrlsub {\isadigit{2}}{\isacharparenright}\ \isanewline
\ \ \ \ \ \ \ \ {\isasymor}\ {\isacharparenleft}findEdge\ w\ {\isacharparenleft}set{\isacharunderscore}to{\isacharunderscore}list\ {\isacharparenleft}parcs\ G{\isacharparenright}{\isacharparenright}\ k{\isacharparenright}\ {\isacharequal}\ {\isacharparenleft}v\isactrlsub {\isadigit{2}}{\isacharcomma}v\isactrlsub {\isadigit{1}}{\isacharparenright}{\isachardoublequoteclose}%
\isadelimproof
\endisadelimproof
\isatagproof
\endisatagproof
{\isafoldproof}%
\isadelimproof
\endisadelimproof
\isadelimproof
\endisadelimproof
\isatagproof
\endisatagproof
{\isafoldproof}%
\isadelimproof
\endisadelimproof
\isadelimproof
\endisadelimproof
\isatagproof
\endisatagproof
{\isafoldproof}%
\isadelimproof
\endisadelimproof
\begin{isamarkuptext}%
We translate the notion of a labelling function $L^i(v)$ (see Definition \ref{def:Labelling}) into Isabelle.
Function \isa{getL\ G\ w}, in short for get label, returns the length of the longest strictly decreasing
path starting at vertex $v$. In contrast to Definition \ref{def:Labelling} subgraphs are treated here implicitly. Intuitively,
this can be seen as adding edges to an empty graph in order of their weight.%
\end{isamarkuptext}\isamarkuptrue%
\isacommand{fun}\isamarkupfalse%
\ getL\ {\isacharcolon}{\isacharcolon}\ {\isachardoublequoteopen}{\isacharparenleft}{\isacharprime}a{\isacharcolon}{\isacharcolon}linorder{\isacharparenright}\ pair{\isacharunderscore}pre{\isacharunderscore}digraph\ {\isasymRightarrow}\ {\isacharparenleft}{\isacharprime}a{\isasymtimes}{\isacharprime}a{\isacharparenright}\ weight{\isacharunderscore}fun\ \isanewline
\ \ \ \ \ \ \ \ \ \ \ \ \ {\isasymRightarrow}\ nat\ {\isasymRightarrow}\ {\isacharprime}a\ {\isasymRightarrow}\ nat{\isachardoublequoteclose}\ \isakeyword{where}\isanewline
{\isachardoublequoteopen}getL\ g\ w\ {\isadigit{0}}\ v\ {\isacharequal}\ {\isadigit{0}}{\isachardoublequoteclose}\ {\isacharbar}\ \ \isanewline
{\isachardoublequoteopen}getL\ g\ w\ {\isacharparenleft}Suc\ i{\isacharparenright}\ v\ {\isacharequal}\ {\isacharparenleft}let\ {\isacharparenleft}v\isactrlsub {\isadigit{1}}{\isacharcomma}v\isactrlsub {\isadigit{2}}{\isacharparenright}\ {\isacharequal}\ {\isacharparenleft}findEdge\ w\ {\isacharparenleft}set{\isacharunderscore}to{\isacharunderscore}list\ {\isacharparenleft}arcs\ g{\isacharparenright}{\isacharparenright}\ {\isacharparenleft}Suc\ i{\isacharparenright}{\isacharparenright}\ in\ \isanewline
\ \ \ \ \ \ \ \ \ \ {\isacharparenleft}if\ v\ {\isacharequal}\ v\isactrlsub {\isadigit{1}}\ then\ max\ {\isacharparenleft}{\isacharparenleft}getL\ g\ w\ i\ v\isactrlsub {\isadigit{2}}{\isacharparenright}{\isacharplus}{\isadigit{1}}{\isacharparenright}\ {\isacharparenleft}getL\ g\ w\ i\ v{\isacharparenright}\ else\ \isanewline
\ \ \ \ \ \ \ \ \ \ {\isacharparenleft}if\ v\ {\isacharequal}\ v\isactrlsub {\isadigit{2}}\ then\ max\ {\isacharparenleft}{\isacharparenleft}getL\ g\ w\ i\ v\isactrlsub {\isadigit{1}}{\isacharparenright}{\isacharplus}{\isadigit{1}}{\isacharparenright}\ {\isacharparenleft}getL\ g\ w\ i\ v{\isacharparenright}\ else\ getL\ g\ w\ i\ v{\isacharparenright}{\isacharparenright}{\isacharparenright}{\isachardoublequoteclose}%
\begin{isamarkuptext}%
To add all edges to the graph, set $i=|E|$. Recall that \isa{card\ {\isacharparenleft}parcs\ g{\isacharparenright}} $= 2*|E|$, 
as every edge appears twice. 
Then, iterate over all vertices and give back the
maximum length which is found by using \isa{getL\ G\ w}. Since \isa{getL\ G\ w} can also be used to get a longest 
strictly increasing trail ending at vertex $v$ the algorithm is not restricted to strictly decreasing trails.%
\end{isamarkuptext}\isamarkuptrue%
\isacommand{definition}\isamarkupfalse%
\ getLongestTrail\ {\isacharcolon}{\isacharcolon}\ \isanewline
{\isachardoublequoteopen}{\isacharparenleft}{\isacharprime}a{\isacharcolon}{\isacharcolon}linorder{\isacharparenright}\ pair{\isacharunderscore}pre{\isacharunderscore}digraph\ {\isasymRightarrow}\ {\isacharparenleft}{\isacharprime}a{\isasymtimes}{\isacharprime}a{\isacharparenright}\ weight{\isacharunderscore}fun\ {\isasymRightarrow}\ nat{\isachardoublequoteclose}\ \isakeyword{where}\isanewline
{\isachardoublequoteopen}getLongestTrail\ g\ w\ {\isacharequal}\ \isanewline
Max\ {\isacharparenleft}set\ {\isacharbrackleft}{\isacharparenleft}getL\ g\ w\ {\isacharparenleft}card\ {\isacharparenleft}parcs\ g{\isacharparenright}\ div\ {\isadigit{2}}{\isacharparenright}\ v{\isacharparenright}\ {\isachardot}\ v\ {\isacharless}{\isacharminus}\ sorted{\isacharunderscore}list{\isacharunderscore}of{\isacharunderscore}set\ {\isacharparenleft}pverts\ g{\isacharparenright}{\isacharbrackright}{\isacharparenright}{\isachardoublequoteclose}%
\begin{isamarkuptext}%
Exporting the algorithm into Haskell code results in a fully verified program to find a longest
strictly decreasing or strictly increasing trail.%
\end{isamarkuptext}\isamarkuptrue%
\isacommand{export{\isacharunderscore}code}\isamarkupfalse%
\ getLongestTrail\ \isakeyword{in}\ Haskell\ \isakeyword{module{\isacharunderscore}name}\ LongestTrail%
\isadelimproof
\endisadelimproof
\isatagproof
\endisatagproof
{\isafoldproof}%
\isadelimproof
\endisadelimproof
\isadelimproof
\endisadelimproof
\isatagproof
\endisatagproof
{\isafoldproof}%
\isadelimproof
\endisadelimproof
\isadelimproof
\endisadelimproof
\isatagproof
\endisatagproof
{\isafoldproof}%
\isadelimproof
\endisadelimproof
\isadelimproof
\endisadelimproof
\isatagproof
\endisatagproof
{\isafoldproof}%
\isadelimproof
\endisadelimproof
\begin{isamarkuptext}%
Using an induction proof and extensive case distinction, the correctness of Algorithm \ref{algo:FindLongestTrail} 
is then shown in our formalization, by proving the following theorem:%
\end{isamarkuptext}\isamarkuptrue%
\isacommand{theorem}\isamarkupfalse%
{\isacharparenleft}\isakeyword{in}\ distinct{\isacharunderscore}weighted{\isacharunderscore}pair{\isacharunderscore}graph{\isacharparenright}\ correctness{\isacharcolon}\isanewline
\ \ \isakeyword{assumes}\ {\isachardoublequoteopen}{\isasymexists}\ v\ {\isasymin}\ {\isacharparenleft}pverts\ G{\isacharparenright}{\isachardot}\ getL\ G\ w\ {\isacharparenleft}q\ div\ {\isadigit{2}}{\isacharparenright}\ v\ {\isacharequal}\ k{\isachardoublequoteclose}\isanewline
\ \ \isakeyword{shows}\ {\isachardoublequoteopen}{\isasymexists}\ xs{\isachardot}\ decTrail\ G\ w\ xs\ {\isasymand}\ length\ xs\ {\isacharequal}\ k{\isachardoublequoteclose}%
\isadelimproof
\endisadelimproof
\isatagproof
\endisatagproof
{\isafoldproof}%
\isadelimproof
\endisadelimproof
\isadelimproof
\endisadelimproof
\isatagproof
\endisatagproof
{\isafoldproof}%
\isadelimproof
\endisadelimproof
\isadelimdocument
\endisadelimdocument
\isatagdocument
\isamarkupsubsection{Minimum Length of Ordered Trails%
}
\isamarkuptrue%
\endisatagdocument
{\isafolddocument}%
\isadelimdocument
\endisadelimdocument
\isadelimproof
\endisadelimproof
\isatagproof
\endisatagproof
{\isafoldproof}%
\isadelimproof
\endisadelimproof
\begin{isamarkuptext}%
\label{section:minLength}
The algorithm introduced in Section \ref{section:computeLongestTrail} is already useful on its own. Additionally, it can be
used to verify the lower bound on the minimum length of a strictly decreasing trail $P_d(w,G) \ge 2 \cdot \lfloor \frac{q}{n} \rfloor$.

To this end, Lemma 1 from Section \ref{section:symbolicProof} is translated into Isabelle as the lemma
\isa{minimal{\isacharunderscore}increase{\isacharunderscore}one{\isacharunderscore}step}. The proof is 
similar to its counterpart, also using a case distinction. Lemma 2 is subsequently proved, here
named \isa{minimal{\isacharunderscore}increase{\isacharunderscore}total}.%
\end{isamarkuptext}\isamarkuptrue%
\isacommand{lemma}\isamarkupfalse%
{\isacharparenleft}\isakeyword{in}\ distinct{\isacharunderscore}weighted{\isacharunderscore}pair{\isacharunderscore}graph{\isacharparenright}\ minimal{\isacharunderscore}increase{\isacharunderscore}one{\isacharunderscore}step{\isacharcolon}\isanewline
\ \ \isakeyword{assumes}\ {\isachardoublequoteopen}k\ {\isacharplus}\ {\isadigit{1}}\ {\isasymin}\ W{\isachardoublequoteclose}\isanewline
\ \ \isakeyword{shows}\ \isanewline
\ \ \ \ {\isachardoublequoteopen}{\isacharparenleft}{\isasymSum}\ v\ {\isasymin}\ pverts\ G{\isachardot}\ getL\ G\ w\ {\isacharparenleft}k{\isacharplus}{\isadigit{1}}{\isacharparenright}\ v{\isacharparenright}\ {\isasymge}\ {\isacharparenleft}{\isasymSum}\ v\ {\isasymin}\ pverts\ G{\isachardot}\ getL\ G\ w\ k\ v{\isacharparenright}\ {\isacharplus}\ {\isadigit{2}}{\isachardoublequoteclose}\ \isanewline
\isadelimproof
\endisadelimproof
\isatagproof
\endisatagproof
{\isafoldproof}%
\isadelimproof
\endisadelimproof
\isadelimproof
\endisadelimproof
\isatagproof
\endisatagproof
{\isafoldproof}%
\isadelimproof
\isanewline
\endisadelimproof
\isacommand{lemma}\isamarkupfalse%
{\isacharparenleft}\isakeyword{in}\ distinct{\isacharunderscore}weighted{\isacharunderscore}pair{\isacharunderscore}graph{\isacharparenright}\ minimal{\isacharunderscore}increase{\isacharunderscore}total{\isacharcolon}\isanewline
\ \ \isakeyword{shows}\ {\isachardoublequoteopen}{\isacharparenleft}{\isasymSum}\ v\ {\isasymin}\ pverts\ G{\isachardot}\ getL\ G\ w\ {\isacharparenleft}q\ div\ {\isadigit{2}}{\isacharparenright}\ v{\isacharparenright}\ {\isasymge}\ q{\isachardoublequoteclose}%
\isadelimproof
\endisadelimproof
\isatagproof
\endisatagproof
{\isafoldproof}%
\isadelimproof
\endisadelimproof
\isadelimproof
\endisadelimproof
\isatagproof
\endisatagproof
{\isafoldproof}%
\isadelimproof
\endisadelimproof
\begin{isamarkuptext}%
From \isa{minimal{\isacharunderscore}increase{\isacharunderscore}total} we have that that the sum of all labels after $q$ div $2$ steps is 
greater than $q$. Now assume that all labels are smaller than $q$ div $n$. Because we have $n$ vertices, this
leads to a contradiction, which proves \isa{algo{\isacharunderscore}result{\isacharunderscore}min}.%
\end{isamarkuptext}\isamarkuptrue%
\isacommand{lemma}\isamarkupfalse%
{\isacharparenleft}\isakeyword{in}\ distinct{\isacharunderscore}weighted{\isacharunderscore}pair{\isacharunderscore}graph{\isacharparenright}\ algo{\isacharunderscore}result{\isacharunderscore}min{\isacharcolon}\ \isanewline
\ \ \isakeyword{shows}\ {\isachardoublequoteopen}{\isacharparenleft}{\isasymexists}\ v\ {\isasymin}\ pverts\ G{\isachardot}\ getL\ G\ w\ {\isacharparenleft}q\ div\ {\isadigit{2}}{\isacharparenright}\ v\ {\isasymge}\ q\ div\ n{\isacharparenright}{\isachardoublequoteclose}%
\isadelimproof
\endisadelimproof
\isatagproof
\endisatagproof
{\isafoldproof}%
\isadelimproof
\endisadelimproof
\begin{isamarkuptext}%
Finally, using lemma \isa{algo{\isacharunderscore}result{\isacharunderscore}min} together with the \isa{correctness} theorem 
of section \ref{section:computeLongestTrail}, we prove the lower bound of $2\cdot\lfloor \frac{q}{n} \rfloor$ over the length 
of a longest strictly decreasing trail. This general approach could also be used to extend our
formalization and prove existence of other trails. For example, assume that some restrictions on the graph 
give raise to the existence of a trail of length $m \ge 2\cdot\lfloor \frac{q}{n} \rfloor$. Then, it is
only necessary to show that our algorithm can find this trail.%
\end{isamarkuptext}\isamarkuptrue%
\isacommand{theorem}\isamarkupfalse%
{\isacharparenleft}\isakeyword{in}\ distinct{\isacharunderscore}weighted{\isacharunderscore}pair{\isacharunderscore}graph{\isacharparenright}\ dec{\isacharunderscore}trail{\isacharunderscore}exists{\isacharcolon}\ \isanewline
\ \ \isakeyword{shows}\ {\isachardoublequoteopen}{\isasymexists}\ es{\isachardot}\ decTrail\ G\ w\ es\ {\isasymand}\ length\ es\ {\isacharequal}\ q\ div\ n{\isachardoublequoteclose}%
\isadelimproof
\endisadelimproof
\isatagproof
\endisatagproof
{\isafoldproof}%
\isadelimproof
\endisadelimproof
\begin{isamarkuptext}%
\end{isamarkuptext}\isamarkuptrue%
\isacommand{theorem}\isamarkupfalse%
{\isacharparenleft}\isakeyword{in}\ distinct{\isacharunderscore}weighted{\isacharunderscore}pair{\isacharunderscore}graph{\isacharparenright}\ inc{\isacharunderscore}trail{\isacharunderscore}exists{\isacharcolon}\ \isanewline
\ \ \isakeyword{shows}\ {\isachardoublequoteopen}{\isasymexists}\ es{\isachardot}\ incTrail\ G\ w\ es\ {\isasymand}\ length\ es\ {\isacharequal}\ q\ div\ n{\isachardoublequoteclose}%
\isadelimproof
\endisadelimproof
\isatagproof
\endisatagproof
{\isafoldproof}%
\isadelimproof
\endisadelimproof
\begin{isamarkuptext}%
Corollary 1 is translated into \isa{dec{\isacharunderscore}trail{\isacharunderscore}exists{\isacharunderscore}complete}. The proof first argues
that the number of edges is $n\cdot(n-1)$ by restricting its domain as done already in Section \ref{section:localeSurjective}.%
\end{isamarkuptext}\isamarkuptrue%
\isadelimproof
\endisadelimproof
\isatagproof
\endisatagproof
{\isafoldproof}%
\isadelimproof
\endisadelimproof
\isacommand{lemma}\isamarkupfalse%
{\isacharparenleft}\isakeyword{in}\ distinct{\isacharunderscore}weighted{\isacharunderscore}pair{\isacharunderscore}graph{\isacharparenright}\ dec{\isacharunderscore}trail{\isacharunderscore}exists{\isacharunderscore}complete{\isacharcolon}\ \isanewline
\ \ \isakeyword{assumes}\ {\isachardoublequoteopen}complete{\isacharunderscore}digraph\ n\ G{\isachardoublequoteclose}\ \isanewline
\ \ \isakeyword{shows}\ {\isachardoublequoteopen}{\isasymexists}\ es{\isachardot}\ decTrail\ G\ w\ es\ {\isasymand}\ length\ es\ {\isacharequal}\ n{\isacharminus}{\isadigit{1}}{\isachardoublequoteclose}%
\isadelimproof
\endisadelimproof
\isatagproof
\endisatagproof
{\isafoldproof}%
\isadelimproof
\endisadelimproof
\isadelimdocument
\endisadelimdocument
\isatagdocument
\isamarkupsubsection{Example Graph $K_4$%
}
\isamarkuptrue%
\endisatagdocument
{\isafolddocument}%
\isadelimdocument
\endisadelimdocument
\begin{isamarkuptext}%
We return to the example graph from Figure \ref{example:K4} and show that our results from 
Sections \ref{section:trails}-\ref{section:minLength} can be used to prove existence of trails of length $k$, in particular
$k = 3$ in $K_4$. Defining the graph and the 
weight function separately, we use natural numbers as vertices.%
\end{isamarkuptext}\isamarkuptrue%
\isacommand{abbreviation}\isamarkupfalse%
\ ExampleGraph{\isacharcolon}{\isacharcolon}\ {\isachardoublequoteopen}nat\ pair{\isacharunderscore}pre{\isacharunderscore}digraph{\isachardoublequoteclose}\ \isakeyword{where}\ \isanewline
{\isachardoublequoteopen}ExampleGraph\ {\isasymequiv}\ {\isacharparenleft}{\isacharbar}\ \isanewline
\ \ pverts\ {\isacharequal}\ {\isacharbraceleft}{\isadigit{1}}{\isacharcomma}{\isadigit{2}}{\isacharcomma}{\isadigit{3}}{\isacharcomma}{\isacharparenleft}{\isadigit{4}}{\isacharcolon}{\isacharcolon}nat{\isacharparenright}{\isacharbraceright}{\isacharcomma}\ \isanewline
\ \ parcs\ {\isacharequal}\ {\isacharbraceleft}{\isacharparenleft}v\isactrlsub {\isadigit{1}}{\isacharcomma}v\isactrlsub {\isadigit{2}}{\isacharparenright}{\isachardot}\ v\isactrlsub {\isadigit{1}}\ {\isasymin}\ {\isacharbraceleft}{\isadigit{1}}{\isacharcomma}{\isadigit{2}}{\isacharcomma}{\isadigit{3}}{\isacharcomma}{\isacharparenleft}{\isadigit{4}}{\isacharcolon}{\isacharcolon}nat{\isacharparenright}{\isacharbraceright}\ {\isasymand}\ v\isactrlsub {\isadigit{2}}\ {\isasymin}\ {\isacharbraceleft}{\isadigit{1}}{\isacharcomma}{\isadigit{2}}{\isacharcomma}{\isadigit{3}}{\isacharcomma}{\isacharparenleft}{\isadigit{4}}{\isacharcolon}{\isacharcolon}nat{\isacharparenright}{\isacharbraceright}\ {\isasymand}\ v\isactrlsub {\isadigit{1}}\ {\isasymnoteq}\ v\isactrlsub {\isadigit{2}}{\isacharbraceright}\ \isanewline
{\isacharbar}{\isacharparenright}{\isachardoublequoteclose}\ \isanewline
\isanewline
\isacommand{abbreviation}\isamarkupfalse%
\ ExampleGraphWeightFunction\ {\isacharcolon}{\isacharcolon}\ {\isachardoublequoteopen}{\isacharparenleft}nat{\isasymtimes}nat{\isacharparenright}\ weight{\isacharunderscore}fun{\isachardoublequoteclose}\ \isakeyword{where}\ \isanewline
{\isachardoublequoteopen}ExampleGraphWeightFunction\ {\isasymequiv}\ {\isacharparenleft}{\isasymlambda}{\isacharparenleft}v\isactrlsub {\isadigit{1}}{\isacharcomma}v\isactrlsub {\isadigit{2}}{\isacharparenright}{\isachardot}\ \isanewline
\ \ {\isacharparenleft}if\ {\isacharparenleft}v\isactrlsub {\isadigit{1}}\ {\isacharequal}\ {\isadigit{1}}\ {\isasymand}\ v\isactrlsub {\isadigit{2}}\ {\isacharequal}\ {\isadigit{2}}{\isacharparenright}\ {\isasymor}\ {\isacharparenleft}v\isactrlsub {\isadigit{1}}\ {\isacharequal}\ {\isadigit{2}}\ {\isasymand}\ v\isactrlsub {\isadigit{2}}\ {\isacharequal}\ {\isadigit{1}}{\isacharparenright}\ then\ {\isadigit{1}}\ else\isanewline
\ \ {\isacharparenleft}if\ {\isacharparenleft}v\isactrlsub {\isadigit{1}}\ {\isacharequal}\ {\isadigit{1}}\ {\isasymand}\ v\isactrlsub {\isadigit{2}}\ {\isacharequal}\ {\isadigit{3}}{\isacharparenright}\ {\isasymor}\ {\isacharparenleft}v\isactrlsub {\isadigit{1}}\ {\isacharequal}\ {\isadigit{3}}\ {\isasymand}\ v\isactrlsub {\isadigit{2}}\ {\isacharequal}\ {\isadigit{1}}{\isacharparenright}\ then\ {\isadigit{3}}\ else\isanewline
\ \ {\isacharparenleft}if\ {\isacharparenleft}v\isactrlsub {\isadigit{1}}\ {\isacharequal}\ {\isadigit{1}}\ {\isasymand}\ v\isactrlsub {\isadigit{2}}\ {\isacharequal}\ {\isadigit{4}}{\isacharparenright}\ {\isasymor}\ {\isacharparenleft}v\isactrlsub {\isadigit{1}}\ {\isacharequal}\ {\isadigit{4}}\ {\isasymand}\ v\isactrlsub {\isadigit{2}}\ {\isacharequal}\ {\isadigit{1}}{\isacharparenright}\ then\ {\isadigit{6}}\ else\isanewline
\ \ {\isacharparenleft}if\ {\isacharparenleft}v\isactrlsub {\isadigit{1}}\ {\isacharequal}\ {\isadigit{2}}\ {\isasymand}\ v\isactrlsub {\isadigit{2}}\ {\isacharequal}\ {\isadigit{3}}{\isacharparenright}\ {\isasymor}\ {\isacharparenleft}v\isactrlsub {\isadigit{1}}\ {\isacharequal}\ {\isadigit{3}}\ {\isasymand}\ v\isactrlsub {\isadigit{2}}\ {\isacharequal}\ {\isadigit{2}}{\isacharparenright}\ then\ {\isadigit{5}}\ else\ \isanewline
\ \ {\isacharparenleft}if\ {\isacharparenleft}v\isactrlsub {\isadigit{1}}\ {\isacharequal}\ {\isadigit{2}}\ {\isasymand}\ v\isactrlsub {\isadigit{2}}\ {\isacharequal}\ {\isadigit{4}}{\isacharparenright}\ {\isasymor}\ {\isacharparenleft}v\isactrlsub {\isadigit{1}}\ {\isacharequal}\ {\isadigit{4}}\ {\isasymand}\ v\isactrlsub {\isadigit{2}}\ {\isacharequal}\ {\isadigit{2}}{\isacharparenright}\ then\ {\isadigit{4}}\ else\isanewline
\ \ {\isacharparenleft}if\ {\isacharparenleft}v\isactrlsub {\isadigit{1}}\ {\isacharequal}\ {\isadigit{3}}\ {\isasymand}\ v\isactrlsub {\isadigit{2}}\ {\isacharequal}\ {\isadigit{4}}{\isacharparenright}\ {\isasymor}\ {\isacharparenleft}v\isactrlsub {\isadigit{1}}\ {\isacharequal}\ {\isadigit{4}}\ {\isasymand}\ v\isactrlsub {\isadigit{2}}\ {\isacharequal}\ {\isadigit{3}}{\isacharparenright}\ then\ {\isadigit{2}}\ else\ {\isadigit{0}}{\isacharparenright}{\isacharparenright}{\isacharparenright}{\isacharparenright}{\isacharparenright}{\isacharparenright}{\isacharparenright}{\isachardoublequoteclose}%
\isadelimproof
\endisadelimproof
\isatagproof
\endisatagproof
{\isafoldproof}%
\isadelimproof
\endisadelimproof
\isadelimproof
\endisadelimproof
\isatagproof
\endisatagproof
{\isafoldproof}%
\isadelimproof
\endisadelimproof
\isadelimproof
\endisadelimproof
\isatagproof
\endisatagproof
{\isafoldproof}%
\isadelimproof
\endisadelimproof
\begin{isamarkuptext}%
We show that the graph $K_4$ of Figure \ref{example:K4} satisfies the conditions that were
imposed in 
\isa{distinct{\isacharunderscore}weighted{\isacharunderscore}pair{\isacharunderscore}graph} and its parent locale, including for example no self loops 
and distinctiveness. Of course there is still some effort required for this. However, it is necessary
to manually construct trails or list all possible weight distributions. Additionally, instead of 
$q!$ statements there are at most $\frac{3q}{2}$ statements needed.%
\end{isamarkuptext}\isamarkuptrue%
\isacommand{interpretation}\isamarkupfalse%
\ example{\isacharcolon}\ \isanewline
\ \ distinct{\isacharunderscore}weighted{\isacharunderscore}pair{\isacharunderscore}graph\ ExampleGraph\ ExampleGraphWeightFunction%
\isadelimproof
\endisadelimproof
\isatagproof
\endisatagproof
{\isafoldproof}%
\isadelimproof
\endisadelimproof
\begin{isamarkuptext}%
Now it is an easy task to prove that there is a trail of length 3. We only add the fact that
\isa{ExampleGraph} is a \isa{distinct{\isacharunderscore}weighted{\isacharunderscore}pair{\isacharunderscore}graph} and lemma \isa{dec{\isacharunderscore}trail{\isacharunderscore}exists}.%
\end{isamarkuptext}\isamarkuptrue%
\isacommand{lemma}\isamarkupfalse%
\ ExampleGraph{\isacharunderscore}decTrail{\isacharcolon}\isanewline
\ \ {\isachardoublequoteopen}{\isasymexists}\ xs{\isachardot}\ decTrail\ ExampleGraph\ ExampleGraphWeightFunction\ xs\ {\isasymand}\ length\ xs\ {\isacharequal}\ {\isadigit{3}}{\isachardoublequoteclose}%
\isadelimproof
\endisadelimproof
\isatagproof
\endisatagproof
{\isafoldproof}%
\isadelimproof
\endisadelimproof
\isadelimdocument
\endisadelimdocument
\isatagdocument
\isamarkupsection{Discussion and Related Work%
}
\isamarkuptrue%
\endisatagdocument
{\isafolddocument}%
\isadelimdocument
\endisadelimdocument
\begin{isamarkuptext}%
Our theory \isa{Ordered{\isacharunderscore}Trail} builds on top of the \isa{Graph{\isacharunderscore}Theory} library presented in \cite{Graph_Theory-AFP}.
However, this library does not formalize strictly ordered trails, nor the special weighted graphs we 
introduced in the locale \isa{distinct{\isacharminus}weighted{\isacharminus}pair{\isacharminus}graph}. 
Furthermore, our formalization extends \cite{Graph_Theory-AFP} with definitions on strictly decreasing and 
increasing trails and provides many basic lemmas on them. Some of the main challenges in this context 
were the reasoning on the surjectivity of the weight function as well the correctness proof of the algorithm.

Our formalization can be easily extended and could therefore serve as a basis for further work in this field.
The definitions \isa{incTrail} and \isa{decTrail} and the respective properties that are proven in 
Section \ref{section:trails} are the key to many other variants of trail properties. 

Graham et al.~\cite{graham1973increasing} also showed upper bounds for trails in complete graphs by decomposing
them into either into cycles or 1-factors. We are currently working on formalizing and certifying the result 
that 

$$
f_d(n)= f_i(n) = 
\begin{cases}
n &  \text{if } n \in \{3,5\},\\
n-1 & \text{otherwise},\\
\end{cases}
$$

\noindent that is, for complete graphs with $n=3$ or $n=5$ vertices there always has to be a trail of length at least $n$ whereas 
for any other number $n$ of vertices there only has to be a trail of length $n - 1$. Therefore, the lower bound that
we showed in this paper is equal to the exact length with exception of two special cases.  
We believe that formalizing this result would be a valuable extension to the theory \isa{Ordered{\isacharunderscore}Trail}.

Another direction for further investigation are monotone paths. 
Graham et al. \cite{graham1973increasing} show that in a complete graph with $n$ vertices there has to be an increasing path of length 
at least $\frac{1}{2}(\sqrt{4n-3}-1)$ and at most $\frac{3n}{4}$. 
The upper bound was afterwards improved by Calderbank, Chung and Sturtevant \cite{calderbank1984increasing}, 
Milans \cite{milans2015monotone} and Buci{\'c} et al. \cite{bucic2018nearly}. 

Recently, other classes of graphs have been considered, e.g., trees and planar graphs \cite{roditty2001monotone},
on random edge-ordering \cite{yuster2001large} or on hypercubes \cite{de2015increasing}.%
\end{isamarkuptext}\isamarkuptrue%
\isadelimdocument
\endisadelimdocument
\isatagdocument
\isamarkupsection{Conclusion%
}
\isamarkuptrue%
\endisatagdocument
{\isafolddocument}%
\isadelimdocument
\endisadelimdocument
\begin{isamarkuptext}%
In this work we formalized strictly increasing and strictly decreasing trails in the proof assistant Isabelle/HOL. 
Furthermore, we showed correctness of an algorithm to find such trails. We provided a verified algorithm and program to compute monotone trails. 
We used this algorithm to
prove the result that every graph with $n$ vertices and $q$ edges has a strictly decreasing trail of length at least
$2\cdot\lfloor\frac{q}{n}\rfloor$. For further work we plan to show that this is a tight bound for every $n$ except for $n = 3$ and $5$. 

Our results are built on the already existing Isabelle \isa{Graph{\isacharunderscore}theory} from the Archive of Formal Proofs. 
Thus, our results can be used by any theory using graphs that are specified as in this library.
Therefore, our theory is highly reusable and might be the basis for further work in this field.%
\end{isamarkuptext}\isamarkuptrue%
\isadelimtheory
\endisadelimtheory
\isatagtheory
\endisatagtheory
{\isafoldtheory}%
\isadelimtheory
\endisadelimtheory
\end{isabellebody}%

\bigskip\noindent\textbf{Acknowledgements}. We thank Prof.~Byron Cook (AWS) for interesting discussions on reasoning challenges with ordered trails. This work was funded by the ERC Starting Grant 2014 SYMCAR
639270, the ERC Proof of Concept Grant 2018 SYMELS 842066, the Wallenberg Academy
Fellowship 2014 TheProSE, the Austrian FWF research project W1255-N23 and P32441,  the Vienna Science and Technology Fund ICT19-065 and the Austrian-Hungarian collaborative project 101\"ou8.

\bibliographystyle{splncs04}
\bibliography{biblio} 

\end{document}